\documentclass{article}
\usepackage{iclr2024_conference,times}
\iclrfinalcopy

\usepackage{amsmath,graphicx,amssymb}
\usepackage[export]{adjustbox}
\usepackage[caption=false]{subfig}
\usepackage{color}
\usepackage{xcolor}
\usepackage{enumitem}
\usepackage{multirow}
\usepackage[page]{appendix}
\usepackage{listings}
\PassOptionsToPackage{hyphens}{url}\usepackage{hyperref}

\hypersetup{colorlinks=true}

\definecolor {CodeBackground}   {rgb}{0.95,  0.95,  0.95}

\newcommand\revise[1]{#1}

\title{Version Control of Speaker Recognition Systems}

\author{%
Quan Wang \qquad 
Ignacio Lopez Moreno \\
Google LLC, USA \\
\href{mailto:quanw@google.com}{\nolinkurl{quanw@google.com}} \qquad
  \href{mailto:elnota@google.com}{\nolinkurl{elnota@google.com}}
}

\fancypagestyle{firstpage}{
  \lhead{\begin{picture}(0,0)\put(0,-3){\includegraphics[width=0.16\linewidth]{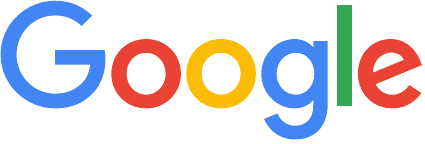}}\end{picture}}
}

\begin{document}

\maketitle
\thispagestyle{firstpage}
\begin{abstract}
This paper discusses one of the most challenging practical engineering problems in speaker recognition systems --- the version control of models and user profiles.
A typical speaker recognition system consists of two stages: the \emph{enrollment stage}, where a profile is generated from user-provided enrollment audio; and the \emph{runtime stage}, where the voice identity of the runtime audio is compared against the stored profiles. As technology advances, the speaker recognition system needs to be updated for better performance. However, if the stored user profiles are not updated accordingly, version mismatch will result in meaningless recognition results. In this paper, we describe different version control strategies for speaker recognition systems that had been carefully studied at Google from years of engineering practice.
These strategies are categorized into three groups according to how they are deployed in the production environment: device-side deployment, server-side deployment, and hybrid deployment.
To compare different strategies with quantitative metrics under various network configurations, we present \texttt{SpeakerVerSim}, an easily-extensible Python-based simulation framework for different server-side deployment strategies of speaker recognition systems.
\end{abstract}

\section{Introduction}
\label{sec:intro}

\subsection{\revise{Speaker recognition systems}}

\emph{Speaker recognition} is the process of recognizing the personal identity of a spoken utterance. Depending on the number of speaker candidates to be recognized, it is often referred to as \emph{speaker verification} (single candidate) or \emph{speaker identification} (multiple candidates). According to the textual content of the spoken utterance being recognized, a speaker recognition  task falls into three categories: \emph{text-dependent} speaker recognition~\cite{hebert2008text}, where the text of the utterance is always the same (\emph{e.g.} a keyword~\cite{chen2014small} or password), or from a very small set of similarly pronounced phrases; \emph{text-prompted} speaker recognition, where the text of the utterance is randomly selected from a predefined large set to prevent spoofing attacks; and \emph{text-independent} speaker recognition~\cite{kinnunen2010overview}, where there is no restriction on the text of the utterance (\emph{e.g.} personalized keyphrases~\cite{rikhye2021personalized}).

The modeling technology for speaker recognition has been an active research field since the 1960s~\cite{kersta1962voiceprint}. With the rapid development of deep learning technologies~\cite{lecun2015deep} in the past decade, the performance of speaker recognition models has greatly improved, especially from long short-term memory (LSTM)~\cite{hochreiter1997long} and attention-based models~\cite{vaswani2017attention,rahman2018attention}.
Large scale commercial speaker recognition systems have emerged in the industry, which can be largely attributed to the popularity of personal voice assistants on smartphones and smart home devices~\cite{pinsky2017tomato}.

\begin{figure}
	\centering
	\includegraphics[width=1\textwidth]{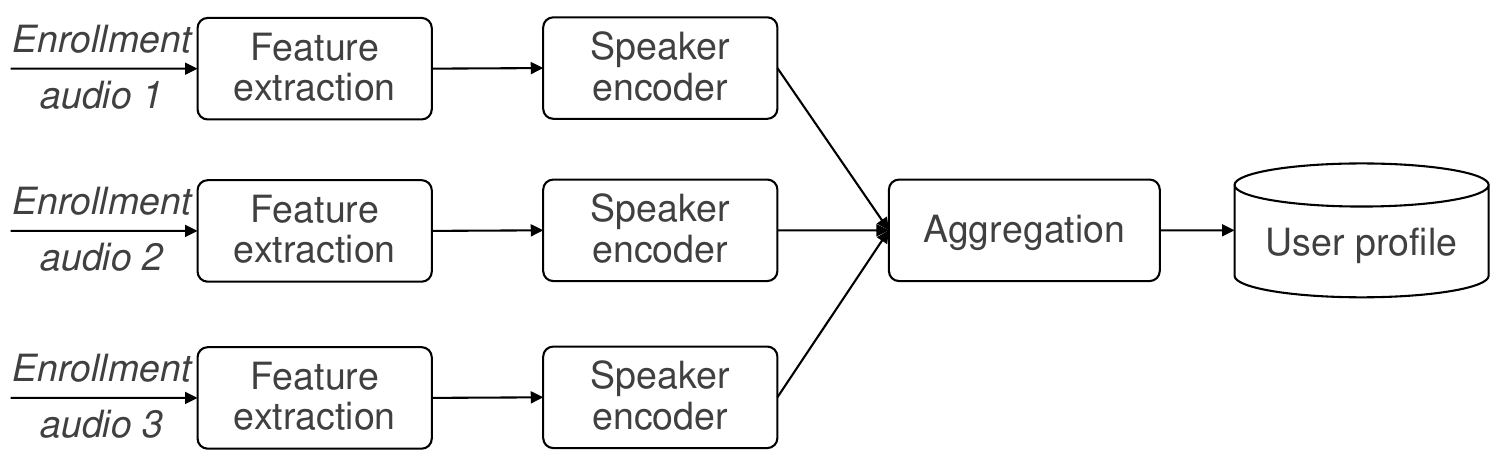}
	\caption{Workflow of the enrollment stage of a speaker recognition system.}
	\label{fig:enrollment_workflow}
\end{figure}

\begin{figure}
	\centering
	\includegraphics[width=1\textwidth]{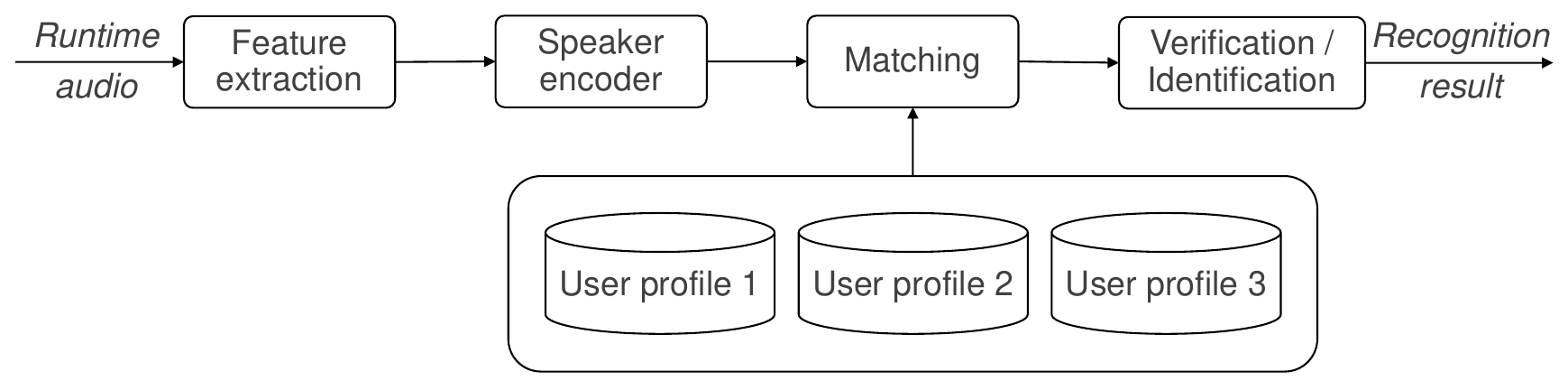}
	\caption{Workflow of the runtime stage of a speaker recognition system.}
	\label{fig:runtime_recognition_workflow}
\end{figure}

Regardless of the number of speaker candidates, the text of utterance, or the specific underlying technology, all speaker recognition systems require two stages of user interaction when deployed to production environment: the \emph{enrollment stage} and the \emph{runtime stage}:
\begin{itemize}
    \item During the enrollment stage, a user provides multiple audio samples to the system, and the system generates a \emph{user profile} to represent the voice characteristics of this user, as shown in Fig.~\ref{fig:enrollment_workflow}.
    \item Once the users have completed the enrollment, the system is ready for runtime recognition, where the voice characteristics of the runtime audio is compared against the enrolled user profiles, as shown in Fig.~\ref{fig:runtime_recognition_workflow}.
\end{itemize}

In both the enrollment stage and the runtime stage, the speaker recognition system needs to extract acoustic features such as perceptual linear prediction (PLP)~\cite{hermansky1990perceptual}, Mel-frequency cepstral coefficients (MFCC)~\cite{davis1980comparison}, power-normalized cepstral coefficients (PNCC)~\cite{kim2016power} or log Mel-filterbank energies (LFBE) from the audio signals. After the acoustic features have been extracted, a speaker encoder model will be used to represent the audio by a speaker embedding, which is usually a fixed-dimension of floating point numbers, such as a Gaussian mixture model (GMM) supervector~\cite{reynolds2000speaker}, speaker factors from joint factor analysis~\cite{kenny2005joint}, an i-vector~\cite{dehak2010front}, or a neural network embedding~\cite{wan2018generalized,li2017deep,snyder2018x}. In the context of this paper, the software that implements feature extraction and speaker encoder will be referred to as the \emph{speech engine}, as they are the most computationally expensive components in the speaker recognition system.

In the enrollment stage, since the system typically requires each user to provide multiple enrollment audio, there will be an additional aggregation step that aggregates speaker encoder outputs to a single embedding vector. In most systems, this aggregation step is an averaging operation of multiple speaker embedding vectors. In recent attentive approaches where each speaker embedding can be interpreted as a group of key-value pairs ~\cite{pelecanos2022parameter}, this aggregation step can be implemented as a joining operation of key-value pairs from speaker encoder outputs.

In the runtime stage, the speaker encoder output is matched against enrolled user profiles via similarity scoring approaches. While cosine similarity and Euclidean distance are the most commonly used similarity scoring approaches, people also studied more complicated approaches using trainable covariance matrix~\cite{snyder2016deep}, decision networks~\cite{pelecanos2021dr}, or parameter-free attentive scoring~\cite{pelecanos2022parameter}.

\subsection{The version control problem}
\label{sec:problem_statement}

After a speaker recognition system has been deployed to production environment, we may still want to update the system for many reasons, including:
\begin{enumerate}
    \item Updating the feature extraction component for better performance (\emph{e.g.} using more frequency bands, or adding new signal processors such as automatic gain control~\cite{prabhavalkar2015automatic}).
    \item Updating the underlying speaker encoder technology for better performance (\emph{e.g.} migrating from i-vector model to neural network based model).
    \item Based on the same technology, updating the speaker encoder model to use a different neural network topology, a different loss function during training, a different similarity scoring approach, or different training datasets.
    \item Periodic model retraining to mitigate the ``data drift'' issue~\cite{mallick2022matchmaker}.
    \item Software optimization and refactoring to improve system robustness, scalability and maintainability.
\end{enumerate}

Because of the enrollment stage, speaker recognition is a \textbf{stateful} system --- the recognition result of a runtime audio depends on the output of other audio (\emph{i.e.} the enrollment audio). This is very different from other speech systems such as automatic speech recognition (ASR), speech enhancement, and language recognition, where the systems are typically stateless. 

As a consequence, the user profiles obtained during the enrollment process (Fig.~\ref{fig:enrollment_workflow}) are ``version dependent''. Once the speaker recognition system has been updated to a newer version, existing user profiles can no longer be used.

In the following sections, we will discuss strategies to re-enroll the user profiles based on a new version of the system in a production environment\footnote{Without loss of generality, we will refer to the new version of the system as the new ``model'' in the following sections for simplicity.}. The strategies are different based on the type of deployment. According to where the speech engine runs and where the user profiles are stored, we categorize deployment solutions into three groups:
\begin{enumerate}
    \item \underline{Device-side deployment}: The speech engine runs on user devices, and the user profiles are also stored on user devices.
    \item \underline{Server-side deployment}: The speech engine runs on cloud computing servers, and the user profiles are stored on cloud databases.
    \item \underline{Hybrid deployment}: The speech engine runs on cloud computing servers, but the user profiles are stored on user devices. 
\end{enumerate}

\begin{figure}
	\centering
	\includegraphics[width=1\textwidth]{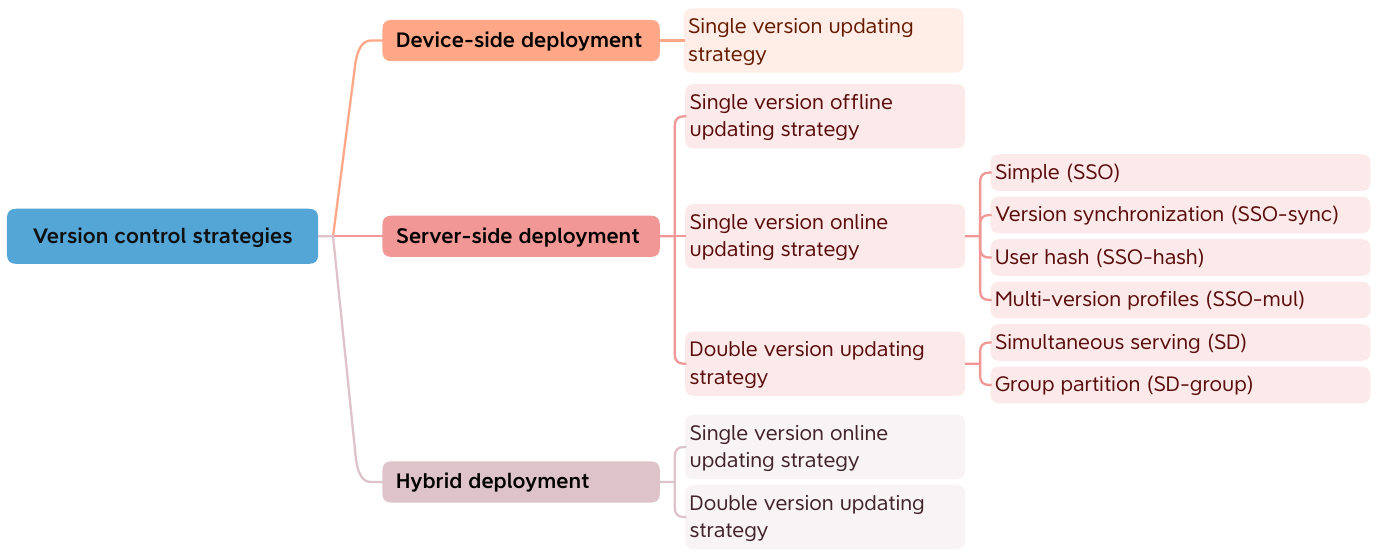}
	\caption{A tree diagram listing all the version control strategies discussed in this paper.}
	\label{fig:strategies-xmind}
\end{figure}

In Fig.~\ref{fig:strategies-xmind}, we list all the version control strategies that will be discussed in the following sections, including the variants.

\subsection{\revise{Contributions}}

The main contribution of this paper is summarized as below:
\begin{enumerate}
    \item As far as we know, this is the first systematic discussion of the version control problem and its solutions for speaker recognition, and more generally, biometric identification systems.
    \item We describe model updating strategies for 3 types of deployment based on the execution of the model and the storage of the user profile, with multiple variants to tackle the version bouncing problem.
    \item We introduce \texttt{SpeakerVerSim}, an easily-extensible Python-based  simulation framework for different version control strategies of speaker recognition systems under different network configurations. This library is open sourced on GitHub\footnote{\url{https://github.com/wq2012/SpeakerVerSim}}.
    \item Using \texttt{SpeakerVerSim} as our testbed, we report a quantitative comparison of different server-side deployment strategies under a typical network configuration.
\end{enumerate}

The rest of this paper is organized as below.
In Section~\ref{sec:related_work}, we briefly review related work in the literature.
In Section~\ref{sec:glossary}, we explain some terms used in the rest of the paper.
In Section~\ref{sec:device}, we introduce version control strategies for device-side deployment. In Section~\ref{sec:server}, we introduce version control strategies for server-side deployment. In Section~\ref{sec:hybrid}, we introduce version control strategies for hybrid deployment. In Section~\ref{sec:exp}, we introduce \texttt{SpeakerVerSim}, a simulation framework for version control strategies of speaker recognition systems, and present simulation results with a quantitative comparison between different version control strategies. Conclusions are drawn in Section~\ref{sec:conclusion} \revise{with a summary of the lessons we learned from this work}.

\section{Related work}
\label{sec:related_work}
\revise{Speaker recognition has been one of the most active research topics in the audio and speech domain. It even has its own special interest group\footnote{\url{https://isca-speech.org/Speaker-and-Language-Characterization-SpLC}} and dedicated workshop\footnote{\url{http://www.speakerodyssey.com}}.
However, the research in the speaker recognition community has been dominantly focusing on the modeling technology. The engineering deployment of speaker recognition systems had been rarely discussed in the literature, including forums like MLSys~\cite{ratner2019mlsys} and Journal of Systems and Software.} As a biometrics problem, speaker recognition shares lots of similarities with other biometric identification systems such as face recognition and fingerprint identification. While researchers have published work related to deployment solutions of such systems~\cite{timur2019deploying,bolme2020face}, we did not find any discussion regarding the version control problem described in Section~\ref{sec:problem_statement}.

\revise{The most relevant work to our paper in the literature is~\cite{limoncelli2014practice}. Chapter 11 of this book introduces strategies for ``updating live services", or in another word, pushing new code to production software environments. Some of these strategies are similar to the server-side deployment strategies we will introduce in Section~\ref{sec:server}. However, biometric identification systems are not as simple as ``code pushes", but have the unique challenge of a separate enrollment stage and the storage of versioned user profiles. These cannot be handled by strategies designed for generic live services.}

In this paper, we offer a comprehensive description of version control strategies that had been studied at Google, with discussions of the advantages and disadvantages of each strategy. Some of the strategies presented in this paper are relatively simple variants of solutions that have already been extensively studied in the parallel computing and computer networks communities~\cite{sanders2019sequential}, but are less familiar to speaker recognition researchers. Although all studies are based on engineering practice with speaker recognition systems at Google, most discussions and conclusions can also apply to other biometric identification systems such as face recognition without loss of generality.

There are many existing powerful frameworks for simulation of very complicated distributed systems, such as SimGrid~\cite{casanova:hal-01017319}, BigSim~\cite{zheng2004bigsim}, and CloudSim~\cite{buyya2009modeling}.
\revise{Since these frameworks are designed for generic network systems, they usually heavily depend on other software libraries, rely on overly complicated configurations, have potential performance inefficiencies, and lack the components (\emph{e.g.} user devices and databases) and metrics (\emph{e.g.} backward version bounce rate) for specific domains. These shortcomings highlight the need for a specialized framework tailored for the version control problem in speaker recognition systems.
In this work, we introduce \texttt{SpeakerVerSim}, an ultra lightweight framework directly built on top of a discrete-event simulation library. Unlike the generic frameworks mentioned above, \texttt{SpeakerVerSim} installs with a single command, only depends on a few small Python libraries, uses a single configuration file where all fields are directly relevant to the problem, and launches the simulation with a single command.}

\section{Glossary}
\label{sec:glossary}
\revise{Here we briefly explain some of the terms we will be using in the rest of this paper in the context of speaker recognition systems:}

\begin{itemize}
    \item \textbf{Speaker embedding}: A vector representing the voice characteristics of a spoken utterance. 
    \item \textbf{Speaker encoder}: The algorithm that generates the speaker embedding from the acoustic features of an utterance.
    \item \textbf{Speech engine}: The software that implements acoustic feature extraction and speaker encoder.
    \item \textbf{User profile}: The aggregated speaker embedding generated from multiple enrollment audio samples provided by the user.
    \item \textbf{Model}: The machine learning model used by the speaker encoder algorithm. In deep learning based approaches, the model is usually a neural network.
    \item \textbf{Frontend}: In server-side and hybrid deployment, the reverse proxy server that dispatches requests from user devices to backend servers. This is also known as the ``master" node in the network.
    \item \textbf{Cloud server}: In server-side and hybrid deployment, the backend server that runs the speech engine. This is also known as the ``worker" node in the network.
    \item \textbf{Database}: In server-side deployment, the backend database that stores enrollment audio and user profiles. 
\end{itemize}

\section{Device-side deployment}
\label{sec:device}

\subsection{Device-side architecture}
In device-side deployment, both the speech engine execution and the user profile storage happen on the user device. The user device could be either smartphones, smart home speakers, or smart security devices. 
For example, a keyword-based text-dependent speaker recognition system could be used to support multiple users on smart home devices~\cite{pinsky2017tomato}; a text-independent speaker recognition system could be used to improve the accuracy of personalized keyphrase detection on smart displays~\cite{rikhye2021personalized}; the enrolled user voice profiles could also be used to improve or personalize other speech systems such as speech enhancement~\cite{Wang2019}, automatic speech recognition~\cite{Wang2020,rikhye2021multiuser,rikhye2022closing,o2023conditional}, voice activity detection~\cite{ding2019personal,ding2022personal}, or text-to-speech synthesis~\cite{jia2018transfer}.
The biggest advantage of device-side deployment is that it does not require any Internet communication with servers at runtime, and all information resides on the same device. This means both enrollment and runtime stages can perform smoothly even when there is no Internet connection.

One big challenge of device-side deployment, however, is the limited computational resources, such as CPU, memory, storage, and power. In most use cases, the user device (\emph{e.g.} a smartphone) needs to perform many other tasks in parallel, thus the resource budget for speaker recognition is usually very limited. There are many approaches to reduce the computational cost of the speaker recognition system, such as model quantization~\cite{alvarez2016efficient,shangguan2019optimizing,ding20224,rybakov20232}, model compression~\cite{nakkiran2015compressing}, model sparsification~\cite{lecun1990optimal,wu2021dynamic} or knowledge distillation~\cite{hinton2015distilling}. Another solution is to implement part of the system on specialized hardware, such as digital signal processors (DSP) or tensor processing units (TPU)~\cite{gupta2021google}.

\subsection{Single version updating strategy}
\begin{figure}
	\centering
	\includegraphics[width=1\textwidth]{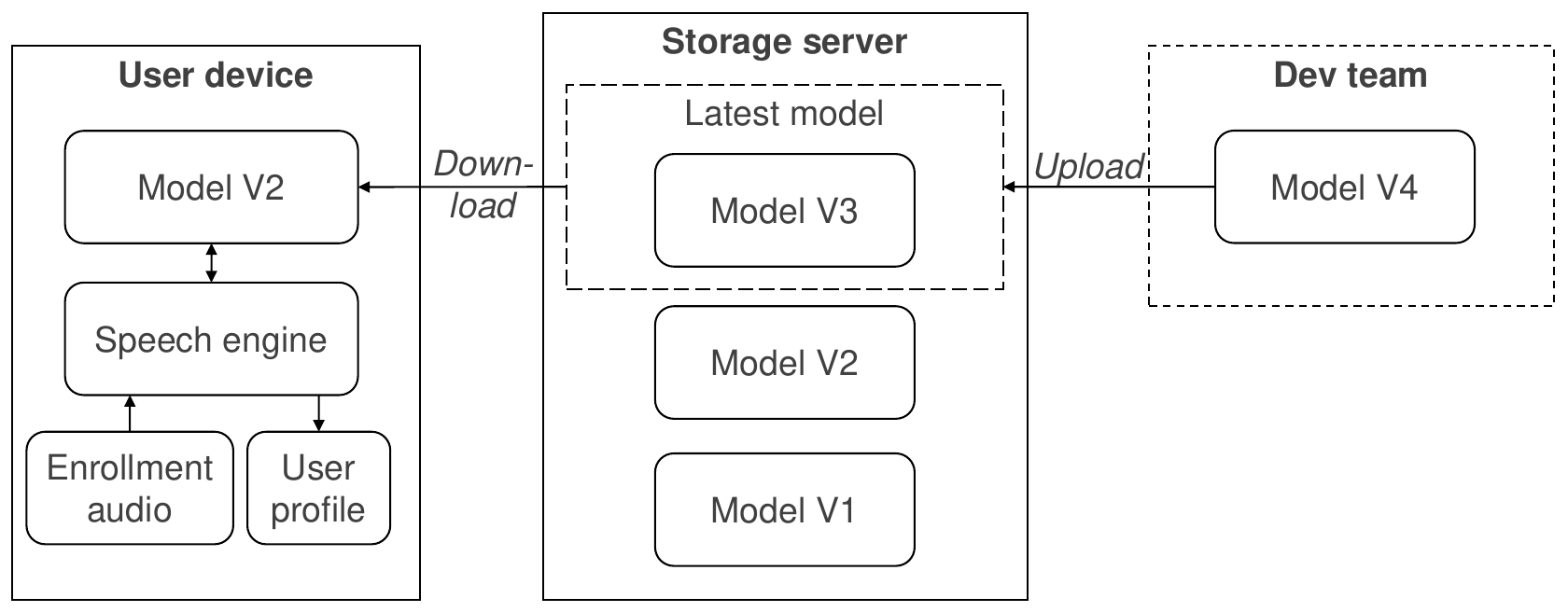}
	\caption{Version control for device-side deployment. The storage server stores all historical models, and provides a shortcut URL for the user device to download the latest model. When the development team uploads a new model, the URL to the latest model will redirect to this new model.}
	\label{fig:on_device_model_update}
\end{figure}

\begin{figure}
	\centering
	\includegraphics[width=0.7\textwidth]{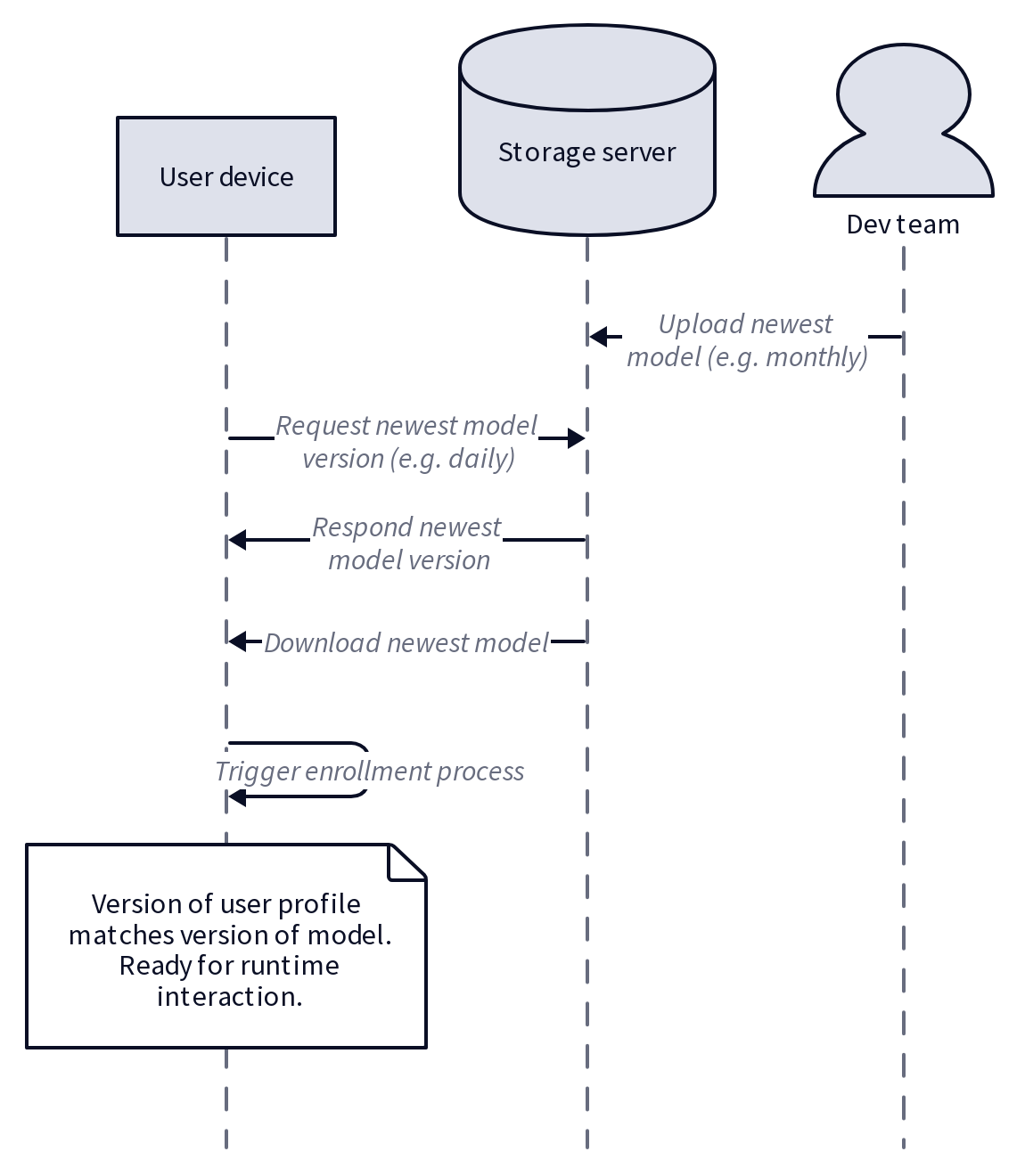}
	\caption{Sequence diagram of the model update process for device-side single version updating strategy.}
	\label{fig:on_device_seqdiagram}
\end{figure}

Version control for device-side deployment is relatively straightforward, as illustrated in Fig.~\ref{fig:on_device_model_update}. The user device only keeps a single model at any specific time to minimize on-device storage. After the enrollment stage, the user's enrollment audio will be stored on the device. When there is a newer version of model available on the model storage server, the user device will download this newer model. Once the download completes, it will immediately trigger a process on the device that uses the newly downloaded model to generate the new version of user profiles based on the enrollment audio. This process guarantees that the version of user profiles stored on the user device always matches the version of the model. \revise{The sequence diagram of this model update process is visualized in Fig.~\ref{fig:on_device_seqdiagram}.}

The only downtime of this single version updating strategy is when the download of the newer model has completed, but the background enrollment process with the newer model has not completed. If a user starts a runtime interaction during this short interval, it will fail. But since the background enrollment process usually completes within a few seconds, this is a very rare case. It can also be avoided by preserving the old model on the device before the enrollment with the newer model completes.

\section{Server-side deployment}
\label{sec:server}

\subsection{Server-side architecture}
\label{sec:server_architecture}
In server-side deployment, both speech engine execution and user profile storage happen on backend servers, which is the opposite of device-side deployment. The biggest advantage of server-side deployment is that the user device only needs to perform very simple operations, such as obtaining the enrollment audio from the user, and communication with the servers. All complicated logic and resource-intensive tasks will be implemented on the servers.
This kind of deployment is usually preferred for very low-end devices such as smart TV remote controllers.

\begin{figure}
	\centering
	\includegraphics[width=1\textwidth]{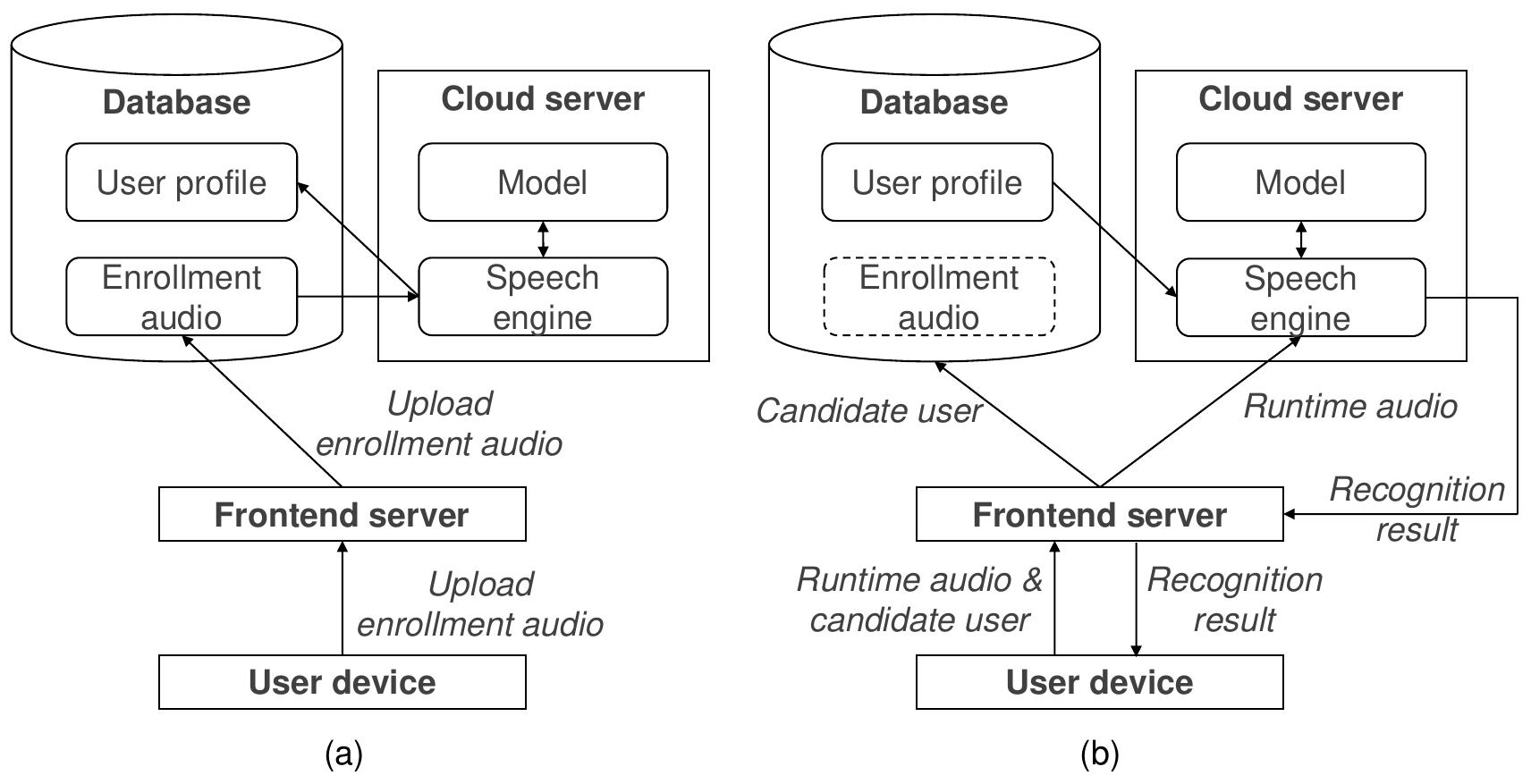}
	\caption{Architecture of server-side deployment of a speaker recognition system. The frontend server is the reverse proxy between the user device and the backend servers; the cloud server performs resource-intensive tasks such as feature extraction and neural network inference; and the database stores each user's enrollment audio and profile. (a) Enrollment stage workflow. (b) Runtime stage workflow.}
	\label{fig:all_server_infra}
\end{figure}

The typical architecture of server-side deployment can be illustrated in Fig.~\ref{fig:all_server_infra}:
\begin{itemize}
    \item During the enrollment stage, the user device first uploads the enrollment audio to the backend database via the frontend reverse proxy server (a ``master" node); next, the speech engine on the cloud computing server (a ``worker" node) generates the user profile based on the enrollment audio; and finally, the user profile will be stored in the backend database. Both the enrollment audio and the user profile are stored together with the user's unique ID.
    \item During the runtime stage, the user device sends the runtime audio together with a set of candidate user IDs to the frontend server; the frontend will fetch the profiles for the candidate users from the backend database, and send them together with the runtime audio to the cloud computing server; finally, the speech engine on the cloud computing server will send the recognition result back to the user device.
\end{itemize}

The request and response schema for enrollment and runtime stages can be roughly described as below:
\lstset{backgroundcolor=\color{CodeBackground},basicstyle=\footnotesize\ttfamily}
\begin{lstlisting}[frame=single]
EnrollmentRequest {
  string user_id;
  vector<Audio> enrollment_audio;
}

EnrollmentResponse {}

RuntimeRequest {
  Audio runtime_audio;
  vector<string> user_id;
}

RuntimeResponse {
  map<string, Result> user_id_to_result;
}
\end{lstlisting}

The problem with the above architecture is obvious: During the runtime stage, if the model on the cloud computing server has been updated to a newer version, it will mismatch the user profiles stored in the database. In the remainder of this section, we will introduce three different version control strategies to handle this problem.

\subsection{Single version offline updating strategy}
\label{sec:server_single_offline}
Among all model updating strategies for server-side deployment, single version offline updating is the simplest one. \revise{This strategy is similar to the ``taking the service down for upgrading" approach for generic live services as introduced in~\cite{limoncelli2014practice}.} Before we update the speaker recognition models in the cloud computing servers, the frontend server will first stop dispatching any new enrollment or runtime requests to the backend. Instead, the frontend will respond to the user device with a special error message, indicating that the backend servers are currently being maintained and updated, and the user device should try again later.

Once the models in the cloud servers have been updated, a background process will be triggered to rerun the enrollment process for all users --- the speech engine will process the enrollment audio for each user, generate a new user profile based on the new model, and replace the existing user profile in the database. Once this large-scale re-enrollment process has been completed, all user profiles in the database will have the same version as the models in the cloud computing servers, and the frontend could resume to accept new enrollment and runtime requests again.

Although this single version offline updating strategy is relatively simple and easy to implement, its disadvantages are also obvious:
\begin{enumerate}
    \item It requires a downtime period of the entire speaker recognition service. If the users are geographically concentrated and the use cases are relatively simple, the updating can be typically scheduled to happen in the local late midnight when we expect very few requests. However, if the users are distributed across multiple time zones, we may expect requests to the service 24 hours a day, thus the downtime will cause significant frustrations to the user experience.
    \item Unlike device-side deployment, where each device only stores the profiles for the owners of the device, in server-side deployment, the database needs to store the profiles of all users. For large-scale applications, the number of users could be huge, thus rerunning enrollment for all users will be a very computationally intensive task. The entire background process of pushing new models to cloud servers plus re-enrolling with the new model may not complete within the scheduled downtime.
\end{enumerate}

\subsection{Single version online updating strategy}
\label{sec:server_single_online}
To avoid the downtime issue in the single version offline updating strategy, an alternative solution is the Server-side Single version Online (abbreviated as \textbf{SSO} in Fig.~\ref{fig:strategies-xmind} and Section~\ref{sec:exp}) updating strategy. In this strategy, we associate each speaker recognition model with a unique \emph{version identifier} string. During the enrollment stage, when we store the user profile in the database, it is stored together with the version identifier of the model that generated it. Then in the runtime stage, when the frontend server receives a new runtime request, it will first check whether the version identifier of the user profile in the database matches the version identifier of the model in the cloud computing server:
\begin{itemize}
    \item If the version identifiers match each other, the frontend server will directly trigger the runtime logic as illustrated in Fig.~\ref{fig:all_server_infra}b.
    \item If the version identifiers do not match, the frontend server will trigger another process to rerun the enrollment  for the user. After the re-enrollment completes, the versions of the user profile and the model are guaranteed to match each other, and the frontend server will trigger the runtime logic.
\end{itemize}

\begin{figure}
	\centering
	\includegraphics[width=0.95\textwidth]{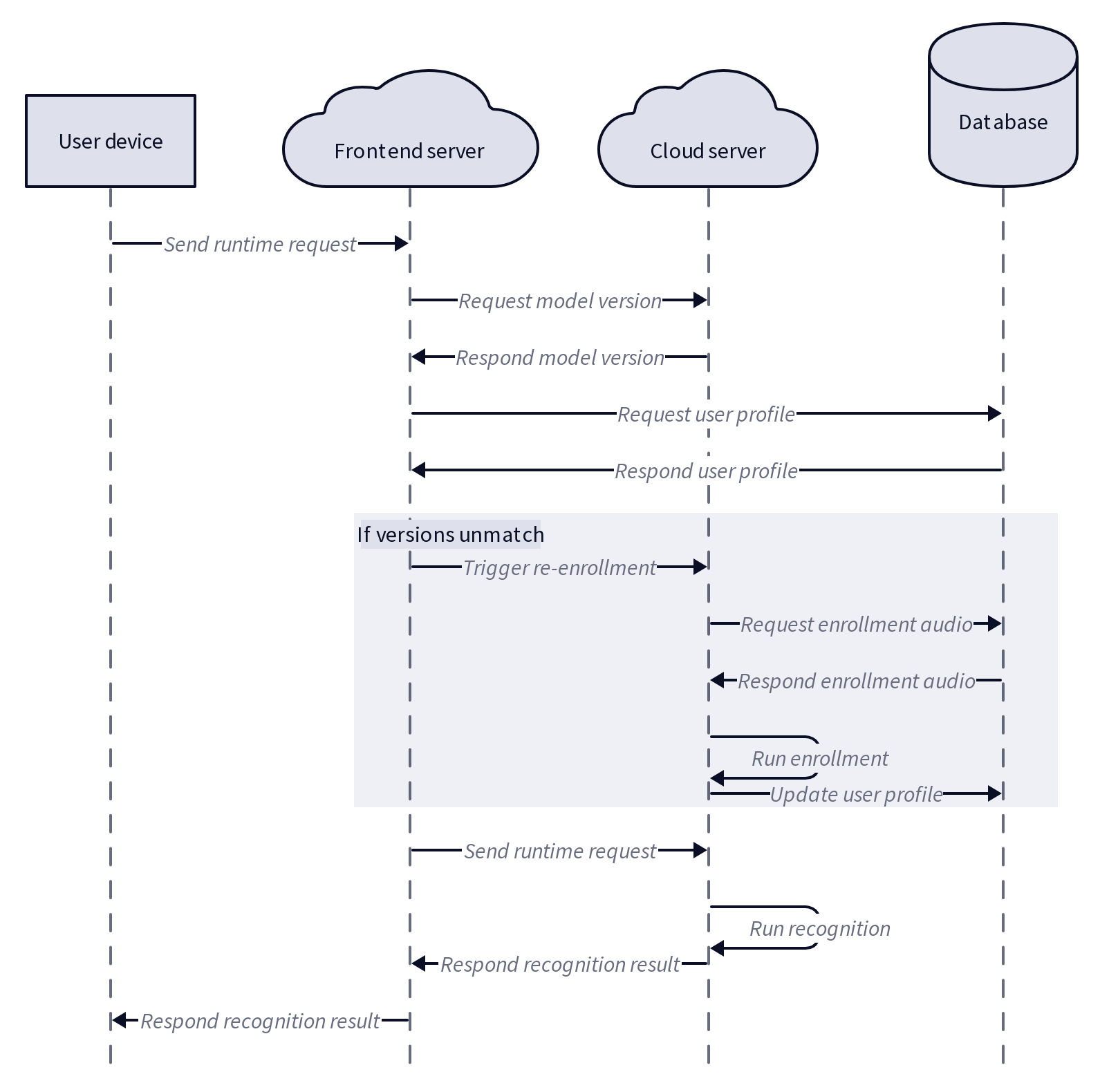}
	\caption{Sequence diagram of a runtime request for server-side single version online (SSO) updating strategy.}
	\label{fig:server_single_seqdiagram}
\end{figure}

As we can see, the single version online updating strategy postpones the re-enrollment process to an on-demand, per-request manner. This guarantees that the speaker recognition service will be available 24 hours a day without downtime. \revise{The sequence diagram of a runtime request in this strategy is visualized in Fig.~\ref{fig:server_single_seqdiagram}.}

However, this strategy also has one disadvantage. Once the model in the cloud computing server has been updated, the next runtime request from each user will always experience increased latency due to the re-enrollment. The significance of the latency increase depends on the efficiency of the re-enrollment process. However, since model updating typically happens every few weeks or months, this increased latency is possibly acceptable for most applications --- it only happens once for each user after each model update.

\begin{figure}
	\centering
	\includegraphics[width=0.7\textwidth]{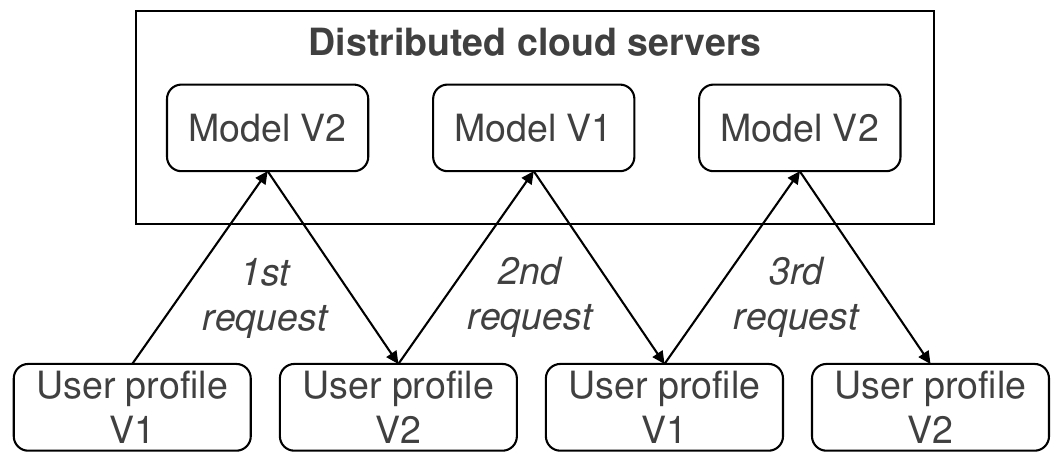}
	\caption{The version bouncing problem for distributed speaker recognition systems.}
	\label{fig:version_bouncing}
\end{figure}

Additionally, for large-scale distributed systems, single version online updating strategy has another challenge known as \textbf{version bouncing}. In a distributed system, there will be multiple cloud computing servers, each serving a copy of the speech engine. When we update the models for the cloud computing servers to a newer version, the update process typically will not finish simultaneously on different machines. This will result in a state that some of the cloud computing servers are serving the new model version, while the other cloud computing servers are still serving the old model version. If a user device sends runtime requests to different servers, the re-enrollment process may happen multiple times, upgrading and degrading the model version in turn, as illustrated in Fig.~\ref{fig:version_bouncing}.

There are several variants of the SSO strategy to avoid the version bouncing problem:
\begin{enumerate}
    \item \textbf{SSO-sync}: The frontend server can periodically send synchronization requests to query the current model version to all cloud computing servers, and maintain a table to record the current model version of each cloud computing server. With this table, if a user profile has been updated, the runtime request will only be dispatched to a cloud computing server with the updated model.
    \item \textbf{SSO-hash}: The frontend server can implement a load balancing algorithm based on the hash value of the user's ID, such that runtime requests for each user are always dispatched to the same cloud computing server. This will guarantee that re-enrollment will only update a user profile from the old version to the new version once.
    \item \textbf{SSO-mul}: Finally, we can store multiple versions of profiles for each user in the database. Once the re-enrollment for a user has completed, we will store both the old version and the new version of this user's profile. For future runtime requests, no matter which version of model is served in the cloud computing server, no re-enrollment will be needed as both versions of profiles are available.
\end{enumerate}

All the above three variants of SSO will effectively reduce the version bouncing rate, thus reducing end-to-end latency and computational cost. Each variant also has its own limitation, which will be further discussed in Section~\ref{sec:exp}.

\subsection{Double version updating strategy}
\label{sec:server_double}

\begin{figure}
	\centering
	\includegraphics[width=1\textwidth]{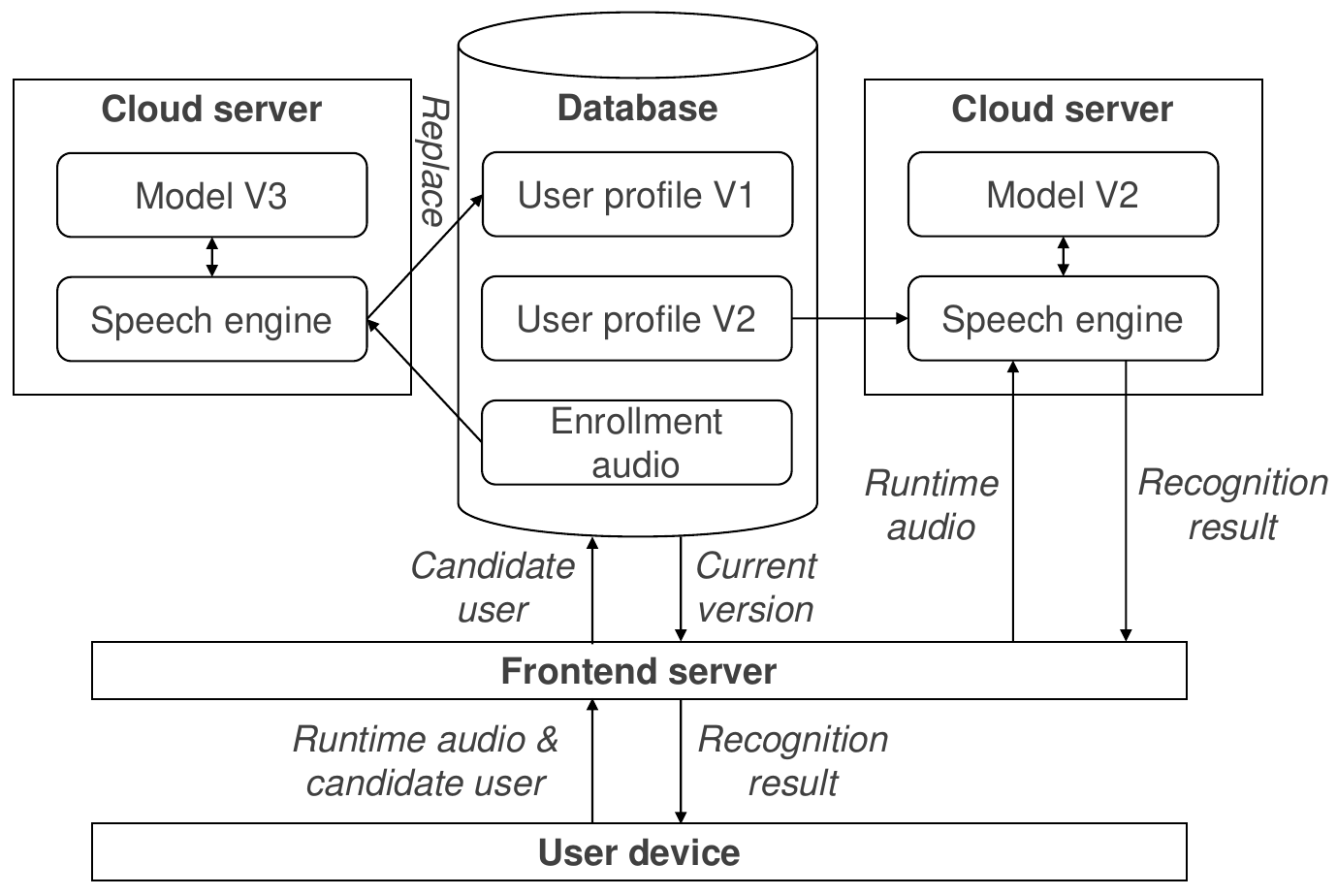}
	\caption{Double version (SD-group) updating strategy for server-side deployment.}
	\label{fig:all_server_double_version_update}
\end{figure}

\begin{figure}
	\centering
	\includegraphics[width=0.95\textwidth]{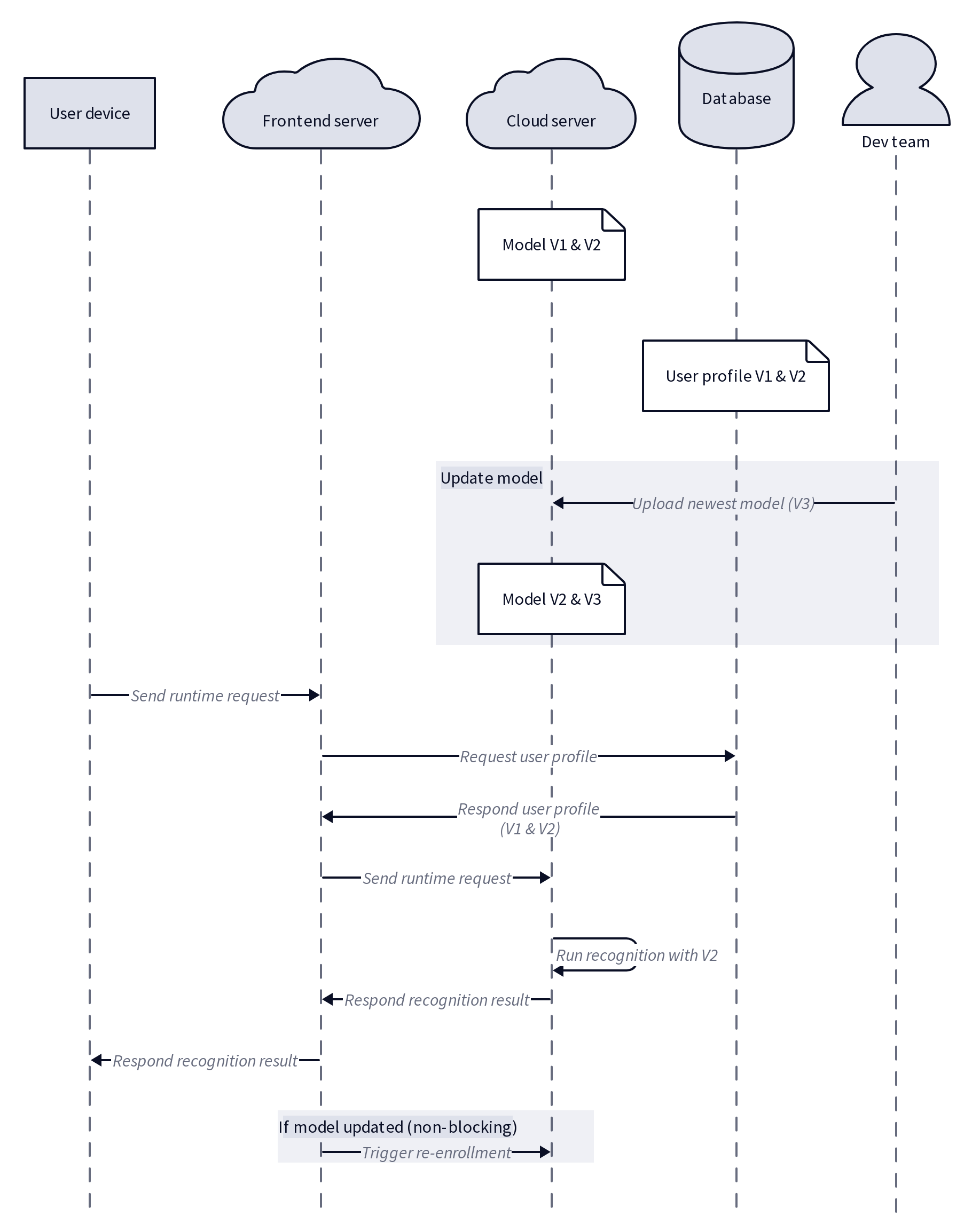}
	\caption{Sequence diagram of a runtime request for server-side double version (SD) updating strategy. Note that no matter whether the models are updated to V2 \& V3 or not, the recognition will be using model V2, as it's the ``newest available model".}
	\label{fig:server_double_seqdiagram}
\end{figure}

As we mentioned before, the single version offline updating strategy requires service downtime for each model update, and the single version online updating strategy will cause increased latency for runtime requests. Here we introduce the Server-side Double version (abbreviated as \textbf{SD} in Fig.~\ref{fig:strategies-xmind} and Section~\ref{sec:exp}) updating strategy, which will overcome these drawbacks.

In the double version updating strategy, we always serve two versions of models in the cloud computing servers at the same time, and always store two versions of user profiles in the database. During the enrollment stage, we always enroll with both models; and during the runtime stage, we use the ``newest available model'', \emph{i.e.} the newest version that is available in both the user profile and the cloud server. The coexistence of two versions guarantees that even if we have updated one model to a newer version, the other model is still available, allowing for a grace period for the user profiles to be updated.
\revise{In this strategy, the sequence diagram of a runtime request is visualized in Fig.~\ref{fig:server_double_seqdiagram}.}

To serve two models, the cloud computing servers can be implemented in two approaches:
\begin{enumerate}
    \item Simultaneous serving (\textbf{SD}): In this implementation, each cloud computing server could serve two models at the same time with separate processes. \revise{It is similar to the ``blue-green deployment" approach for generic live services as introduced in~\cite{limoncelli2014practice}.}
    \item Group partition (\textbf{SD-group}): Alternatively, we could divide the cloud computing servers into two groups, each group serving one model. The group partition is fixed, so the frontend server does not need to periodically synchronize with each of the cloud computing servers. \revise{It is similar to the ``proportional shedding" approach for generic live services as introduced in~\cite{limoncelli2014practice}.}
\end{enumerate}

Assuming the group partition implementation, we use Fig.~\ref{fig:all_server_double_version_update} as an example to illustrate this strategy. Originally, the cloud computing servers were serving model V1 and model V2 simultaneously, and we stored both user profile V1 and V2 in the database. When the development team releases a newer model V3, it will replace the group of cloud computing servers that are still serving the oldest model V1. During this process, the frontend is still handling all enrollment and runtime requests:
\begin{itemize}
    \item Enrollment requests will be dispatched to both cloud computing servers serving model V2 and V3. User profiles for both V2 and V3 will be produced and stored in the database.
    \item For a runtime request, if the user profiles have not been updated (only V1 and V2), the request will be dispatched to a cloud computing server serving model V2. Because user profile V2 is available, the runtime recognition can be performed smoothly without additional latency (as is the case in Fig.~\ref{fig:all_server_double_version_update}). At the same time, the frontend will trigger a re-enrollment process in the background to replace user profile V1 by user profile V3. \revise{Note that this re-enrollment process is non-blocking, so it can be scheduled with lower priority.}
    \item For a runtime request, if the user profiles have already been updated to V2 and V3, the request will be dispatched to a cloud computing server serving model V3 (another case not described in Fig.~\ref{fig:all_server_double_version_update}).
\end{itemize}

As we can see, in the double version updating strategy, while background processes are updating the models on cloud servers to the newer version, and updating user profiles to the newer version, the speaker recognition service will still be always available without additional latency. There will usually be sufficient time to update all user profiles until the next model release. Apparently, this is the most elegant version control solution for server-side deployment, which we will justify in Section~\ref{sec:exp} with quantitative metrics.
The only downside of the double version updating strategy is that its implementation requires additional complexity of the cloud server, such as serving multiple models (increased disk and memory) or grouping.

Since the SD strategy and the SSO-mul strategy described in Section~\ref{sec:server_single_online} both involve storing multiple versions of the user profile in the database, here we clarify their differences:
\begin{enumerate}
    \item Each cloud server in SSO-mul still only serves a single version of the model; cloud servers in SD may serve two versions of the model at the same time.
    \item An re-enrollment request in SSO-mul will block the fulfillment of the runtime request, like other variants of SSO; re-enrollment request in SD runs in the background and is non-blocking for the current runtime request, thus does not introduce any latency.
\end{enumerate}

\section{Hybrid deployment}
\label{sec:hybrid}

\subsection{Hybrid architecture}
In Section~\ref{sec:device} and Section~\ref{sec:server}, we discussed the version control strategies for device-side and server-side deployment. Although device-side deployment is simple and requires no Internet communications, it's not available for many applications where the on-device computational resource budgets are limited. At the same time, storing user profiles on server-side databases may result in privacy concerns for some applications~\cite{de2017europe}.

\begin{figure}
	\centering
	\includegraphics[width=1\textwidth]{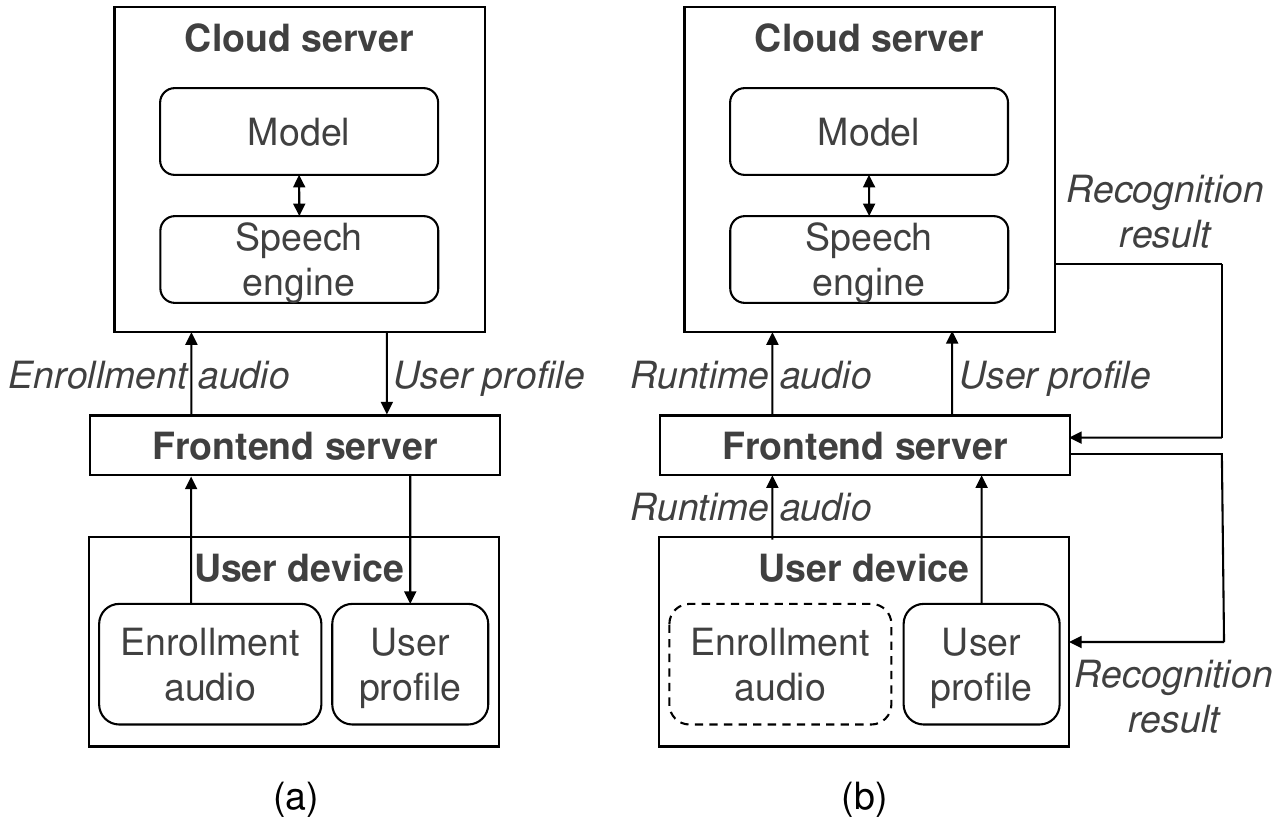}
	\caption{Architecture of hybrid deployment. (a) Enrollment stage workflow. (b) Runtime stage workflow.}
	\label{fig:hybrid_infra}
\end{figure}

An alternative solution is the hybrid deployment, where the speech engine execution happens on cloud computing servers, but the user profiles are stored on user devices, as illustrated in Fig.~\ref{fig:hybrid_infra}:
\begin{itemize}
    \item During the enrollment stage, the user device first sends the enrollment audio to the frontend server; then the speech engine produces the user profile from the enrollment audio; finally, the frontend server will send the user profile back to the user device. Once the enrollment stage completes, the servers will immediately delete the user profile from the memory; the user device is responsible for storing the user profiles.
    \item During the runtime stage, the user device sends the runtime audio together with candidate user profiles to the frontend server; the speech engine will compare the voice identity of the runtime audio against the candidate user profiles; finally, the recognition results will be sent back to the user device. The user profiles are typically encrypted when being stored on the user device and communicated to the servers for security.
\end{itemize}

Similar to Section~\ref{sec:server_architecture}, we provide the rough request and response schema for enrollment and runtime stages of hybrid deployment as below:
\lstset{backgroundcolor=\color{CodeBackground},basicstyle=\footnotesize\ttfamily}
\begin{lstlisting}[frame=single]
EnrollmentRequest {
  string user_id;
  vector<Audio> enrollment_audio;
}

EnrollmentResponse {
  bytes profile;
}

RuntimeRequest {
  Audio runtime_audio;
  map<string, bytes> user_id_to_profile;
}

RuntimeResponse {
  map<string, Result> user_id_to_result;
}
\end{lstlisting}

The hybrid deployment is very similar to the server-side deployment, except that user profiles are stored in the user devices instead of in a backend database. Because all device-server communications can only be initiated by the user device, the servers cannot access the user profiles at arbitrary given time, which poses a new challenge to the hybrid deployment.

\subsection{Single version online updating strategy}
\label{sec:hybrid_single_online}

\begin{figure}
	\centering
	\includegraphics[width=0.8\textwidth]{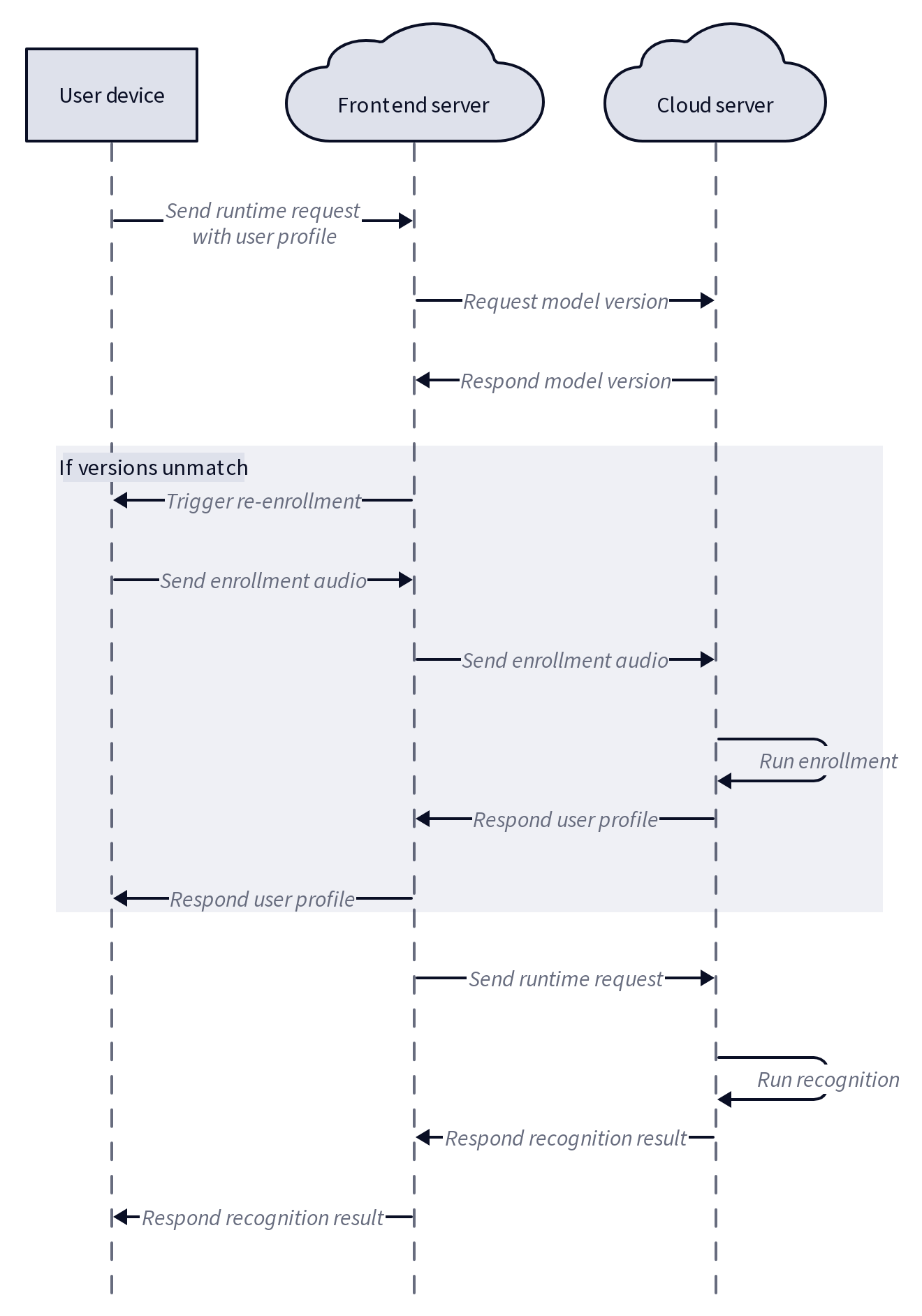}
	\caption{Sequence diagram of a runtime request for hybrid single version online updating strategy. Note that unlike Fig.~\ref{fig:server_single_seqdiagram}, a database is not needed in this strategy.}
	\label{fig:hybrid_single_seqdiagram}
\end{figure}

For hybrid deployment, we could use a single version online updating strategy that is very similar to the strategy we introduced in Section~\ref{sec:server_single_online} for server-side deployment. When the user device sends a runtime request to the frontend server, it will first check whether the user profile version matches the version of the model in the cloud computing server. If the versions do not match, it will trigger the enrollment stage to update the user profile, then perform runtime recognition after the re-enrollment completes. \revise{The sequence diagram of a runtime request for this strategy is visualized in Fig.~\ref{fig:hybrid_single_seqdiagram}.}

Similar to server-side deployment, single version online updating strategy will cause increased latency to the first runtime request for each user after the model has been updated. This could be mitigated by implementing a daily \emph{handshaking} communication between the user device and the server, initiated by the user device. This handshaking communication will simply check whether the version matches between the device and the server; if they mismatch, it will silently trigger the re-enrollment in the background. The handshaking communication could happen at the late midnight in the device's local time zone to minimize user interference.

\subsection{Double version updating strategy}
\label{sec:hybrid_double}

\begin{figure}
	\centering
	\includegraphics[width=0.9\textwidth]{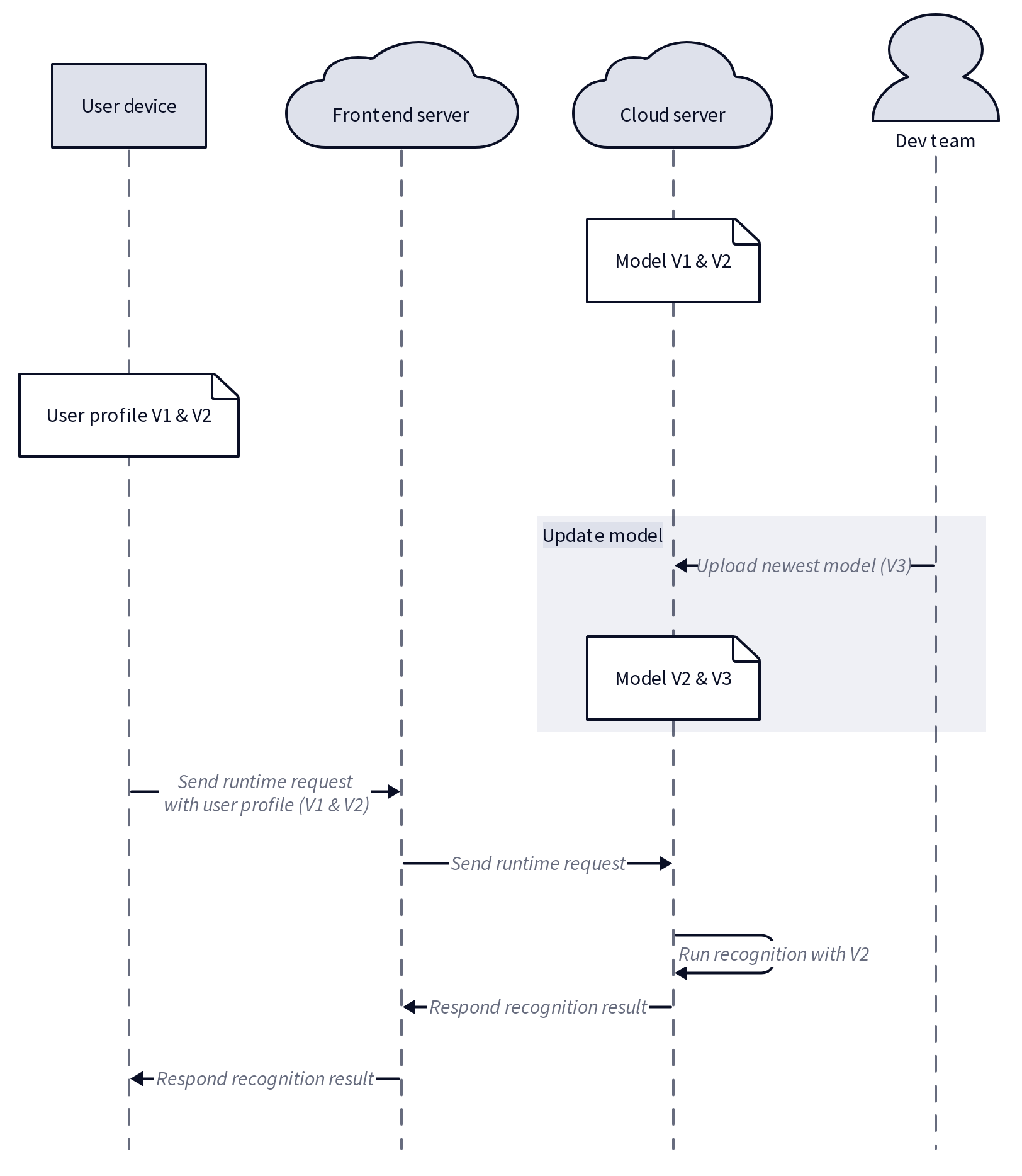}
	\caption{Sequence diagram of a runtime request for hybrid double version updating strategy. Note that unlike Fig.~\ref{fig:server_double_seqdiagram}, a database is not needed in this strategy. Also, re-enrollment is usually not triggered at runtime, but through a periodic handshaking communication initiated by the user device.}
	\label{fig:hybrid_double_seqdiagram}
\end{figure}

For hybrid deployment, we could also use a similar double version updating strategy as the one introduced in Section~\ref{sec:server_double} for server-side deployment. In this strategy, the cloud computing servers always serve two versions of models, and the user devices also always store two versions of user profiles. During the enrollment stage, the server always produces two versions of user profiles and sends them back to the user device. At runtime, even if one server-side model has been updated to a newer version, the other model is still available for those devices whose user profiles have not been updated. \revise{The sequence diagram of a runtime request for this strategy is visualized in Fig.~\ref{fig:hybrid_double_seqdiagram}.} The request and response schema for enrollment and runtime stages would be modified as below:
\lstset{backgroundcolor=\color{CodeBackground},basicstyle=\footnotesize\ttfamily}
\begin{lstlisting}[frame=single]
VersionedProfile {
  map<string, bytes> version_to_profile;
}

EnrollmentRequest {
  string user_id;
  vector<Audio> enrollment_audio;
}

EnrollmentResponse {
  VersionedProfile profile;
}

RuntimeRequest {
  Audio runtime_audio;
  map<string, VersionedProfile> user_id_to_profile;
}

RuntimeResponse {
  map<string, Result> user_id_to_result;
}
\end{lstlisting}

For hybrid deployment, even if we use the double version updating strategy, we still need to make sure that all user devices complete the update within a certain time frame. Otherwise, if some user devices missed two consecutive server-side model updates, both versions of user profiles stored on the device will not be usable. One solution is to implement periodic handshaking communication between the user device and the server, as we mentioned in Section~\ref{sec:hybrid_single_online}.

\section{Experiments}
\label{sec:exp}

\subsection{Simulation testbed}

In order to compare different version control strategies for speaker recognition systems with quantitative metrics, we implemented a library named \texttt{SpeakerVerSim}. \texttt{SpeakerVerSim} is built on top of \texttt{SimPy}, a process-based discrete-event simulation (DES) framework based on standard Python\footnote{\url{https://simpy.readthedocs.io}}.

\revise{In \texttt{SpeakerVerSim}, each machine in the network inherits from the \texttt{Actor} class, including the client, the frontend server, the cloud worker, and the database. All clients inherit from the \texttt{BaseClient} class; all frontend servers inherit from the \texttt{BaseFrontend} class; all cloud workers inherit from the \texttt{BaseWorker} class; and all databases inherit from the \texttt{BaseDatabase} class.
The communication between two machines happens in this way: the sender creates a \texttt{Message} object, and adds it to the receiver's message pool, which is a \texttt{simpy.Store} object; then the receiver pulls from its own message pool with its own process.
During the simulation, metrics are logged in an object of the \texttt{GlobalStats} class.
The entire network system is represented by the \texttt{NetworkSystem} class or its subclass.}

\revise{With the above basic design, each version control strategy is implemented by creating a set of client, frontend server (the ``master" node), cloud servers (the ``worker" node), database, and defining how they interact with each other. 
Once a network system is defined, we can run the simulation and report the communication latency and computational cost of different actors in this network system.}

Using \texttt{SpeakerVerSim} as our testbed, we have implemented 5 different server-side model updating strategies described in Section~\ref{sec:server}: SSO, SSO-sync, SSO-hash, SSO-mul, and SD. We did not implement device-side deployment strategies because they do not involve network communication. We also did not implement hybrid deployment strategies because they are very similar to server-side deployment strategies, thus any conclusions of server-side deployment hold the same for hybrid deployment. \revise{However, we would like to point out that new version control strategies can be easily added by reusing existing components that have already been implemented for the strategies we experimented in this paper.}

\subsection{Configurations}
\label{sec:config}
\texttt{SpeakerVerSim} is a highly configurable framework, where all configurations are passed to the simulation via a single YAML file~\footnote{\url{https://yaml.org}}. In this section, we are going to report simulation results under a specific configuration as our case study. \revise{The values used in this configuration are best effort approximations of empirical data from realistic applications at Google.}

First, we assume all enrollment and runtime audio have a sample rate of 16 kHz, use 16-bit linear PCM audio encoding, and the audio length follows a Gaussian distribution with mean of 5 seconds and standard deviation of 0.5 second, \revise{which is the typical length of a voice query}. We assume the user client device connects to the servers with a 4G network with 5 Mbps bandwidth. The frontend server connects with cloud servers with a high speech network of 1 Gbps bandwidth. The latency of the read and write operations on the backend database is 0.5ms and 10ms, respectively. The real-time factor (RTF) of the speech engine is 0.1. For computational cost of the speaker encoder, we use the numbers reported in~\cite{xia2022turn}, \emph{i.e.} $C_\mathrm{enc}=0.42$ Gflops to process 1 second of audio.

To simulate the model version updating process, we assume that we start to push a new version of the speaker recognition model to all cloud servers at time $t=0$. For each cloud server, the time it takes to complete the model update is an independent and identically distributed (IID) random variable following an exponential distribution, with an expected value of one hour. \revise{The frontend server receives a runtime request from a client every 10 seconds.} For SSO, SSO-mul and SD, the frontend dispatches each request to a random cloud server. For the SSO-sync strategy specifically, we assume the frontend server sends version query requests to all cloud servers every 10 minutes. We run the simulation of the entire network system for 3 hours from $t=0$.

\subsection{Metrics}

Here we introduce the metrics that have been implemented in the \texttt{SpeakerVerSim} library.

\subsubsection{Latency}

The end-to-end latency of fulfilling a client request is the most interesting metric when comparing different version control strategies, as it will directly affect the user experience of the speaker recognition system. The end-to-end latency is defined as the duration from the time when the client sends a runtime request to the frontend server, to the time when the client receives the runtime response containing speaker recognition results from the frontend server. \revise{Assume the $i$th runtime request is sent at time $t_\mathrm{req}^{(i)}$, and the corresponding response is received at time $t_\mathrm{res}^{(i)}$ by the client. The end-to-end latency is computed as below:
\begin{equation}
    \label{eq:latency}
    \tau^{(i)} =t_\mathrm{res}^{(i)} - t_\mathrm{req}^{(i)}.
\end{equation}
}

Here we report the distribution of the end-to-end latency of all requests in one simulation. We also compute the average and maximal end-to-end latency for each simulation, and report its distribution from 100 independent simulations: \revise{
\begin{equation}
    \label{eq:latency_average}
    \tau_\mathrm{avg} =\frac{1}{N} \sum_{i=1}^N \tau^{(i)},
\end{equation}
\begin{equation}
    \label{eq:latency_max}
    \tau_\mathrm{max} = \max_{1 \leq i \leq N} \tau^{(i)},
\end{equation}
where $N$ is the total number of runtime requests fulfilled between $t=0$ and the end of simulation (\emph{i.e.} 3 hours).
}
\subsubsection{Computational cost}

To measure the computational cost in the cloud servers, we report the average floating point operations (flops) being processed in the cloud servers for fulfilling one runtime request. We also compute its distribution from 100
independent simulations.
\revise{For the speaker recognition system, the dominant computational cost is the inference cost of the speaker encoder during the enrollment and recognition. Thus the average flops per request can be computed as:
\begin{equation}
    \label{eq:cost}
    C_\mathrm{avg} = \frac{1}{N} \sum_{i=1}^N {C_\mathrm{enc} \cdot (L^{(i)} + \delta_\mathrm{enroll}^{(i)} \cdot L_\mathrm{enroll}^{(i)})},
\end{equation}
where $C_\mathrm{enc}$ is the inference cost of the speaker encoder to process 1 second of audio (\emph{i.e.} 0.42Gflops), $L^{(i)}$ is the length of the audio from the $i$th runtime request, $\delta_\mathrm{enroll}^{(i)} \in \{0,1\}$ indicates whether the $i$th runtime request triggered blocking re-enrollment, $L_\mathrm{enroll}^{(i)}$ is the total length of enrollment audio corresponding to the speaker from the $i$th runtime request, and $N$ is the total number of runtime requests fulfilled between $t=0$ and the end of simulation.}

\subsubsection{Backward version bounce rate}

In Section~\ref{sec:server_single_online}, we introduced the version bouncing problem. Thus in our simulation, we report the backward version bounce rate, which is defined as the percentage of all runtime requests which triggered re-enrollment using an older model than the version of the current profile, \revise{computed as below:
\begin{equation}
    \label{eq:backward}
    r_\mathrm{backward} = \frac{1}{N} \sum_{i=1}^N \delta_\mathrm{backward}^{(i)},
\end{equation}
where $\delta_\mathrm{backward}^{(i)} \in \{0,1\}$ indicates whether the $i$th runtime request triggered a backward re-enrollment, and $N$ is the total number of runtime requests fulfilled between $t=0$ and the end of simulation.}
Backward version bounce rate is highly correlated with end-to-end latency and computational cost.

\subsubsection{Workload}

We also report the flops being processed by each cloud server at different times during the entire simulation. This will allow us to better understand how the choice of version control strategies will affect our choice of the load balancing algorithm~\cite{sanders2019sequential,ghomi2017load}.

\subsection{Simulation results}

\subsubsection{Impact of version bouncing}

\begin{figure}
	\centering
	\includegraphics[width=0.7\textwidth]{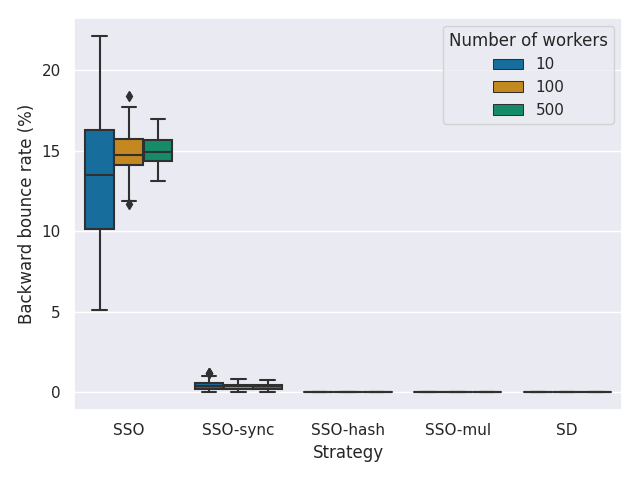}
	\caption{Distribution of backward version bounce rate (Eq.~\ref{eq:backward}) of different version control strategies from 100 independent simulations.}
	\label{fig:backward_bounce_rate}
\end{figure}

\begin{figure}
	\centering
	\includegraphics[width=0.7\textwidth]{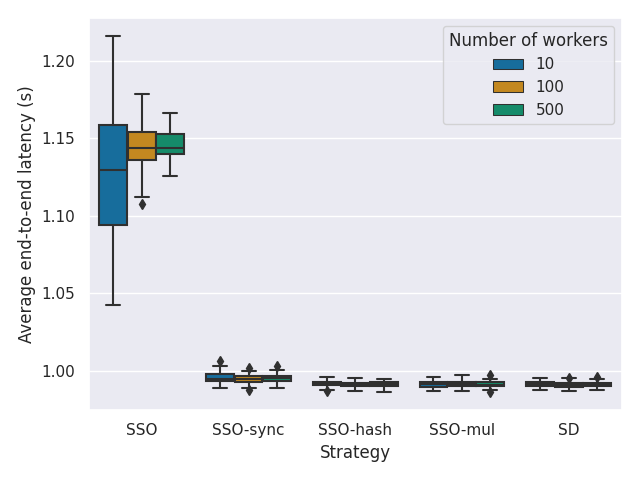}
	\caption{Distribution of average end-to-end latency (Eq.~\ref{eq:latency_average}) of different version control strategies from 100 independent simulations.}
	\label{fig:average_e2e_latency}
\end{figure}

\begin{figure}
	\centering
	\includegraphics[width=0.7\textwidth]{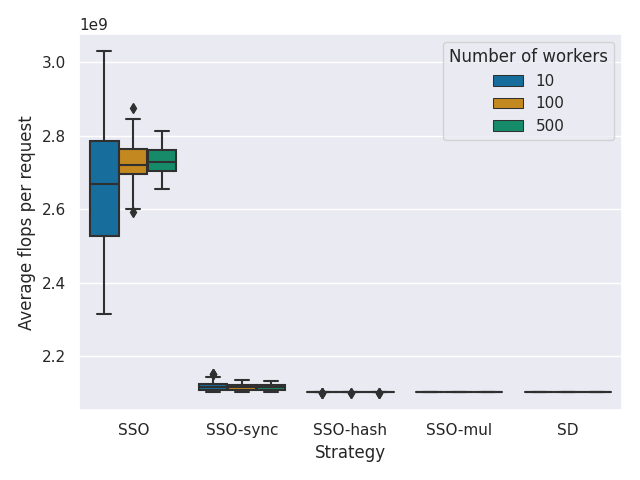}
	\caption{Distribution of average flops per request (Eq.~\ref{eq:cost}) of different version control strategies from 100 independent simulations.}
	\label{fig:average_total_flops}
\end{figure}

For each strategy, we run the simulation with the configuration described in Section~\ref{sec:config} for 100 times. From the 100 independent simulations, the distributions of the backward version bounce rate (Eq.~\ref{eq:backward}), the average end-to-end latency (Eq.~\ref{eq:latency_average}), and the average flops per request (Eq.~\ref{eq:cost}) are visualized in Fig.~\ref{fig:backward_bounce_rate}, Fig.~\ref{fig:average_total_flops}, and Fig.~\ref{fig:average_e2e_latency}, respectively.

From Fig.~\ref{fig:backward_bounce_rate}, we can see that the SSO strategy has the highest backward version bounce rate. SSO-sync effectively reduced the backward version bounce rate. The other strategies, SSO-hash, SSO-mul and SD do not have the backward version bouncing problem at all. These results are well expected according to our discussions in Section~\ref{sec:server_single_online}.

In Fig.~\ref{fig:average_e2e_latency} and Fig.~\ref{fig:average_total_flops}, we can see that SSO has the highest average end-to-end latency and average flops per request. SSO-sync has much smaller values, while SSO-hash, SSO-mul and SD have the smallest. The results in these two figures are highly correlated with the result in Fig.~\ref{fig:backward_bounce_rate}, as the increased  latency and flops are largely due to the cost of re-enrollment from the version bouncing problem.

\begin{figure}
	\centering
	\includegraphics[width=0.7\textwidth]{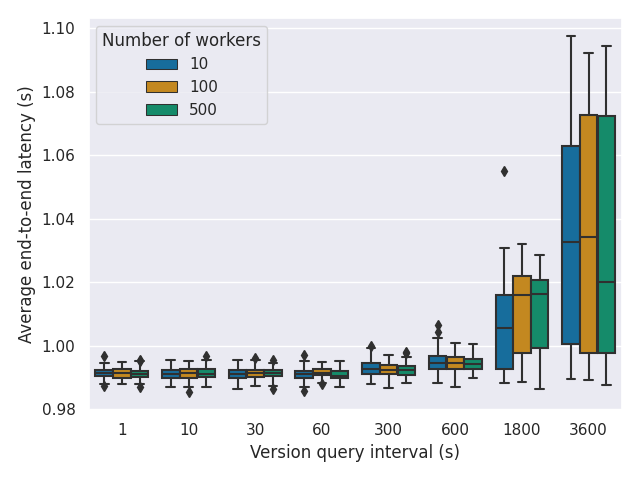}
	\caption{Distribution of average end-to-end latency (Eq.~\ref{eq:latency_average}) for the SSO-sync strategy from 100 independent simulations with different intervals of the version query request.}
	\label{fig:sso_sync_sweep}
\end{figure}

The reason SSO-sync has larger backward version bounce rate than SSO-hash, SSO-mul and SD is due to the potential lagging of the version query between the frontend server and the cloud servers --- a cloud server could have been updated to the newer version, while the table in the frontend server has not updated accordingly. This can be mitigated by making the frontend server send  version query requests more frequently to the cloud servers. In Fig.~\ref{fig:sso_sync_sweep}, we visualize the distribution of average end-to-end latency for the SSO-sync strategy from 100 independent simulations with different intervals of the version query request, ranging from 1 second to 1 hour. As we can see, the average end-to-end latency will be smaller when the version query interval is shorter.

\subsubsection{Maximal end-to-end latency}

\begin{figure}
	\centering
	\includegraphics[width=0.7\textwidth]{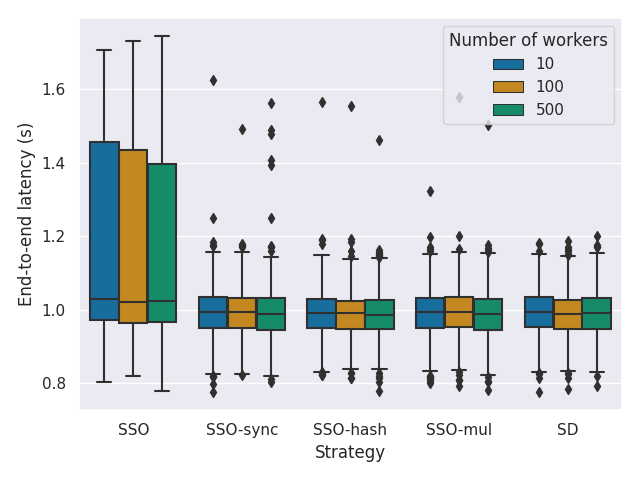}
	\caption{Distribution of the end-to-end latency (Eq.~\ref{eq:latency}) of all requests in a single simulation.}
	\label{fig:e2e_latency}
\end{figure}

\begin{figure}
	\centering
	\includegraphics[width=0.7\textwidth]{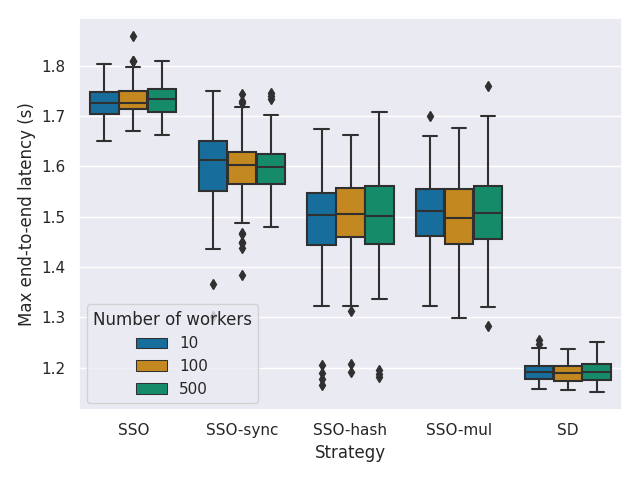}
	\caption{Distribution of maximal end-to-end latency (Eq.~\ref{eq:latency_max}) of different version control strategies from 100 independent simulations.}
	\label{fig:max_e2e_latency}
\end{figure}

While Fig.~\ref{fig:average_e2e_latency} shows the distribution of average end-to-end latency, we are also interested in the maximal end-to-end latency among all requests from one simulation \revise{as computed in Eq.~\ref{eq:latency_max}, as it reflects the worst case from the user perspective}.  In Fig.~\ref{fig:e2e_latency}, we visualize the distribution of the end-to-end latency of all requests from a single simulation. In Fig.~\ref{fig:max_e2e_latency}, we also visualize the distribution of the maximal end-to-end latency from 100 independent simulations.

From Fig.~\ref{fig:e2e_latency}, we can see that the range of end-to-end latency for SSO is pretty big, largely due to re-enrollment for both backward version bouncing and forward version update. For SSO-sync, the range is much smaller due to lower backward version bounce rate.
For SSO-hash and SSO-mul, the range is even smaller with only very few outliers, because there is no backward version bouncing. For SD, there are no outliers since re-enrollment happens in the background. These observations are well reflected in Fig.~\ref{fig:max_e2e_latency}, where we see the highest maximal end-to-end latency for SSO and SSO-sync, smaller maximal end-to-end latency for SSO-hash and SSO-mul, and smallest maximal end-to-end latency for SD.

\subsubsection{Workload of cloud servers}
\begin{figure}
	\centering
	\includegraphics[width=0.9\textwidth]{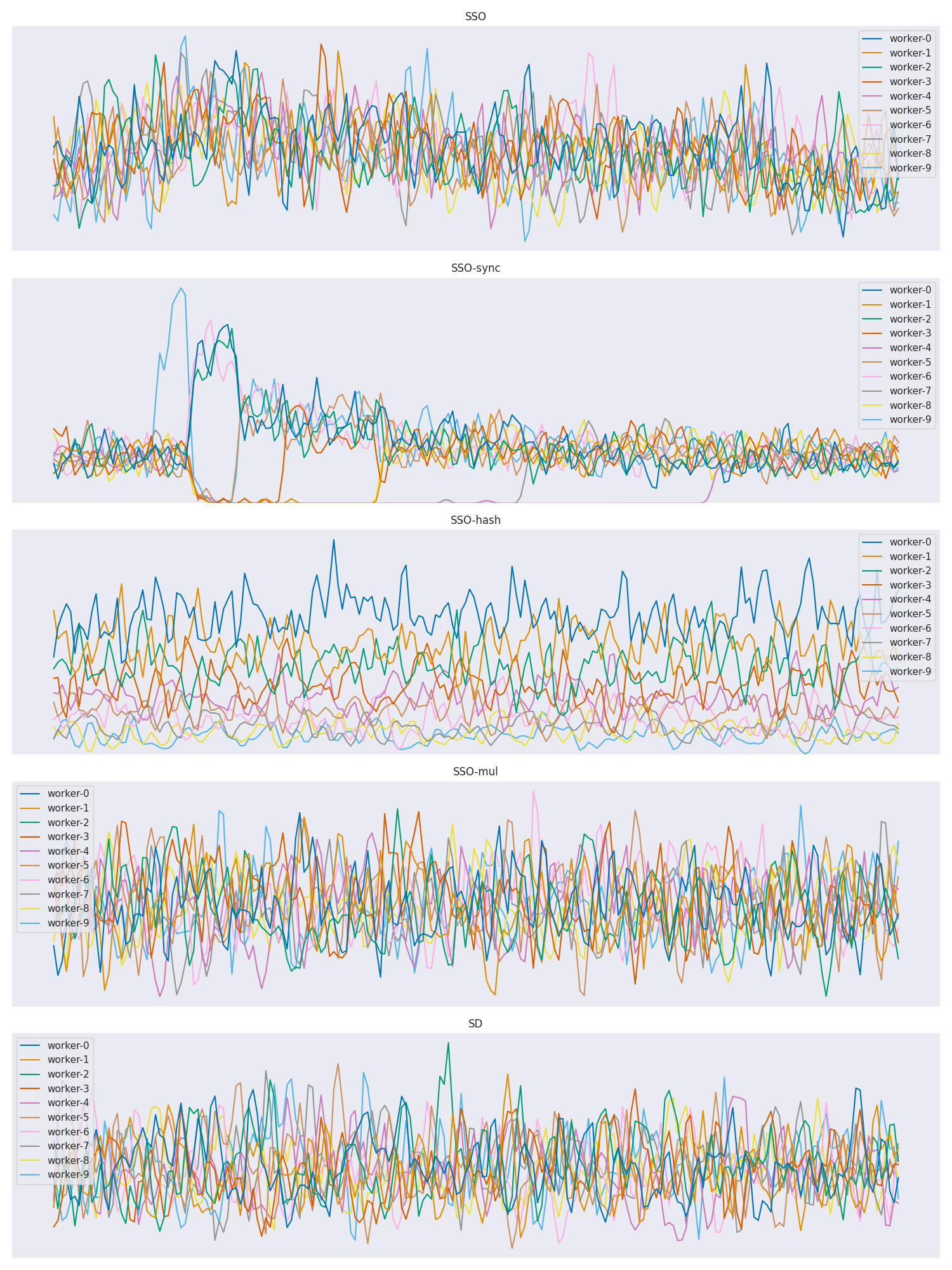}
	\caption{The flops being processed at different times by each backend cloud server during one simulation.}
	\label{fig:workload_10workers_100users}
\end{figure}

To understand how different version control strategies affect the workload of cloud servers, we make a few modifications to the configuration described in Section~\ref{sec:config}. First, we assume the network consists of 10 cloud servers in the backend, and 100 user clients. Different users send runtime requests with different frequencies. The frequencies decrease exponentially when sorted by user. The frontend server receives a runtime request every one second. With this configuration, we use Seaborn~\cite{waskom2021seaborn} to visualize the flops being processed at different times by each backend server during the simulation in
Fig.~\ref{fig:workload_10workers_100users}. \revise{The curves in this figure are smoothed with a Gaussian kernel~\cite{scott2015multivariate}.}

From Fig.~\ref{fig:workload_10workers_100users}, we can see that for SSO, SSO-mul and SD, the workload is uniformly distributed across 10 different cloud servers. This is because for these 3 strategies, the frontend always dispatches the runtime request to a random cloud server.

For SSO-sync, we are seeing a high workload in a few cloud servers (\emph{e.g.} worker-9) after the start of the simulation. As time passes, the workload on some cloud servers (\emph{e.g.} worker-4) will become very low. This is because towards the start of backend version update, only a few cloud servers have updated to the newer version of the model. Once a user has completed re-enrollment with this newer model, its request will always be dispatched to these few cloud servers, causing a spike in their workload. Towards the end of the backend version update, most cloud servers have updated to the newer version of the model, and most users have re-enrolled with the newer model, causing low traffic to those cloud server that are still serving the old model.

For SSO-hash, we are seeing consistently higher workload on some cloud servers (\emph{e.g.} worker-0 and worker-1) than some other cloud servers (\emph{e.g.} worker-8 and worker-9). This is because requests from each user are always dispatched to the same cloud server, and in this simulation, each user sends requests with a very different frequency.

From Section~\ref{sec:server_single_online} and Section~\ref{sec:server_double}, we know that SSO, SSO-mul and SD strategies do not have any assumptions on how the frontend dispatches the requests to cloud servers, which means we could use any load balancing algorithm~\cite{sanders2019sequential} that works best for the network configuration.
SSO-sync and SSO-hash, however, require additional frontend dispatching logic, which is less flexible when a specific load balancing algorithm (\emph{e.g.} dynamic load balancing~\cite{alakeel2010guide,ghomi2017load}) is preferred.
Specifically, from the simulation results above, SSO-sync and SSO-hash may fail to make proper use of cloud computing resources under some configurations (\emph{e.g.} low resource in worker-9 for SSO-sync, or low resource in worker-0 for SSO-hash).

\subsection{\revise{Reproducing the simulation results}}

We have open sourced the \texttt{SpeakerVerSim} library on GitHub as \url{https://github.com/wq2012/SpeakerVerSim}. To run a simulation with this library \revise{and reproduce the results from this paper}, one can simply run a command like:
\lstset{backgroundcolor=\color{CodeBackground},basicstyle=\footnotesize\ttfamily}
\begin{lstlisting}[frame=single]
python run_simulator.py -c my_config.yml -s SSO-sync
\end{lstlisting}

In the example command above, \texttt{my\_config.yml} is a YAML file containing all the configurations of the simulation to be run. The flag \texttt{-s} or \texttt{--strategy} can be used to override the strategy being used in the simulation, \emph{e.g.} \texttt{SSO-sync} in the above command. The \texttt{example\_config.yml} config file provided with this library has a detailed explanation of all the configurable fields.

\section{Conclusions}
\label{sec:conclusion}
In this paper, we introduced the concept of version control in speaker recognition systems. Version control is a common and challenging problem when deploying biometric identification systems such as speaker recognition systems to production environments.
Based on how we execute the speech engine and how we store the user profiles, we categorize speaker recognition deployment solutions into three groups: device-side deployment, server-side deployment, and hybrid deployment. We introduced version control strategies for each type of deployment, and discussed the advantages and disadvantages of each strategy.
To quantitatively compare different version control strategies, we presented \texttt{SpeakerVerSim}, a Python-based framework to simulate version control strategies of speaker recognition systems under various network conditions. In a case study of server-side deployment strategies, we demonstrated that a double version updating strategy is preferred over a single version online updating strategy and its variants based on analysis of backward version bounce rate, maximal end-to-end latency, and workload of the backend cloud servers.

\revise{In addition, we would like to emphasize several lessons learned from our experience with practical development and deployment of speaker recognition systems. First of all, based on computational and legal constraints, the same machine learning model and technology may require different deployment solutions for different products. For example, device-side deployment would be viable for flagship smartphones; server-side deployment is more feasible for smart home speakers; while hybrid deployment is a better alternative to server-side deployment in regions with higher legal risk on biometrics. For each deployment solution, the version control strategy needs to be designed, implemented and evaluated accordingly. Secondly, we have found it extremely valuable to simulate each version control strategy under different configurations before actually implementing it in the production system. Many practical issues, such as the version bouncing problem, are discovered during the simulation, and motivated us to improve our design of the strategy. Thirdly, while there are many existing generic network simulation frameworks, it's usually more expedient to implement a lightweight framework tailored to the problem directly based on discrete-event simulation libraries. This allows for rapid development of the simulation system, easy customization of specialized components and metrics, and avoids overly complicated dependencies and configurations. Lastly, speaker recognition has been a widely studied problem since the 1960s, and has its own special interest group and dedicated workshop in the research community. However, we are surprised to find out that almost no prior work in the literature discusses practical deployment problems for speaker recognition systems such as the version control problem. We would like to challenge the speaker recognition community to explore beyond the modeling approaches, and call for more research and publications on the system design and software engineering aspects of speaker recognition.}

\newpage
\bibliographystyle{IEEEtran}
\bibliography{mybib}

\begin{thebibliography}{10}
\providecommand{\url}[1]{#1}
\csname url@samestyle\endcsname
\providecommand{\newblock}{\relax}
\providecommand{\bibinfo}[2]{#2}
\providecommand{\BIBentrySTDinterwordspacing}{\spaceskip=0pt\relax}
\providecommand{\BIBentryALTinterwordstretchfactor}{4}
\providecommand{\BIBentryALTinterwordspacing}{\spaceskip=\fontdimen2\font plus
\BIBentryALTinterwordstretchfactor\fontdimen3\font minus \fontdimen4\font\relax}
\providecommand{\BIBforeignlanguage}[2]{{%
\expandafter\ifx\csname l@#1\endcsname\relax
\typeout{** WARNING: IEEEtran.bst: No hyphenation pattern has been}%
\typeout{** loaded for the language `#1'. Using the pattern for}%
\typeout{** the default language instead.}%
\else
\language=\csname l@#1\endcsname
\fi
#2}}
\providecommand{\BIBdecl}{\relax}
\BIBdecl

\bibitem{hebert2008text}
M.~H{\'e}bert, ``Text-dependent speaker recognition,'' in \emph{Springer Handbook of Speech Processing}.\hskip 1em plus 0.5em minus 0.4em\relax Springer, 2008, pp. 743--762.

\bibitem{chen2014small}
G.~Chen, C.~Parada, and G.~Heigold, ``Small-footprint keyword spotting using deep neural networks,'' in \emph{International Conference on Acoustics, Speech and Signal Processing (ICASSP)}.\hskip 1em plus 0.5em minus 0.4em\relax IEEE, 2014, pp. 4087--4091.

\bibitem{kinnunen2010overview}
T.~Kinnunen and H.~Li, ``An overview of text-independent speaker recognition: From features to supervectors,'' \emph{Speech Communication}, vol.~52, no.~1, pp. 12--40, 2010.

\bibitem{rikhye2021personalized}
R.~Rikhye, Q.~Wang, Q.~Liang, Y.~He, D.~Zhao, A.~Narayanan, I.~McGraw \emph{et~al.}, ``Personalized keyphrase detection using speaker and environment information,'' in \emph{Proc. Interspeech}, 2021.

\bibitem{kersta1962voiceprint}
L.~G. Kersta, ``Voiceprint identification,'' \emph{The Journal of the Acoustical Society of America}, vol.~34, no. 5\_Supplement, pp. 725--725, 1962.

\bibitem{lecun2015deep}
Y.~LeCun, Y.~Bengio, and G.~Hinton, ``Deep learning,'' \emph{Nature}, vol. 521, no. 7553, pp. 436--444, 2015.

\bibitem{hochreiter1997long}
S.~Hochreiter and J.~Schmidhuber, ``Long short-term memory,'' \emph{Neural computation}, vol.~9, no.~8, pp. 1735--1780, 1997.

\bibitem{vaswani2017attention}
A.~Vaswani, N.~Shazeer, N.~Parmar, J.~Uszkoreit, L.~Jones, A.~N. Gomez, {\L}.~Kaiser, and I.~Polosukhin, ``Attention is all you need,'' \emph{Advances in neural information processing systems}, vol.~30, 2017.

\bibitem{rahman2018attention}
F.~R. rahman Chowdhury, Q.~Wang, I.~L. Moreno, and L.~Wan, ``Attention-based models for text-dependent speaker verification,'' in \emph{International Conference on Acoustics, Speech and Signal Processing (ICASSP)}.\hskip 1em plus 0.5em minus 0.4em\relax IEEE, 2018, pp. 5359--5363.

\bibitem{pinsky2017tomato}
Y.~Pinsky, ``Tomato, tomahto. {Google Home} now supports multiple users,'' \emph{Google Assistant Blog}, 2017.

\bibitem{hermansky1990perceptual}
H.~Hermansky, ``Perceptual linear predictive ({PLP}) analysis of speech,'' \emph{the Journal of the Acoustical Society of America}, vol.~87, no.~4, pp. 1738--1752, 1990.

\bibitem{davis1980comparison}
S.~Davis and P.~Mermelstein, ``Comparison of parametric representations for monosyllabic word recognition in continuously spoken sentences,'' \emph{IEEE Transactions on Acoustics, Speech, and Signal Processing}, vol.~28, no.~4, pp. 357--366, 1980.

\bibitem{kim2016power}
C.~Kim and R.~M. Stern, ``Power-normalized cepstral coefficients ({PNCC}) for robust speech recognition,'' \emph{IEEE/ACM Transactions on Audio, Speech, and Language Processing}, vol.~24, no.~7, pp. 1315--1329, 2016.

\bibitem{reynolds2000speaker}
D.~A. Reynolds, T.~F. Quatieri, and R.~B. Dunn, ``Speaker verification using adapted {Gaussian} mixture models,'' \emph{Digital Signal Processing}, vol.~10, no. 1-3, pp. 19--41, 2000.

\bibitem{kenny2005joint}
P.~Kenny, ``Joint factor analysis of speaker and session variability: Theory and algorithms,'' \emph{CRIM, Montreal,(Report) CRIM-06/08-13}, vol.~14, pp. 28--29, 2005.

\bibitem{dehak2010front}
N.~Dehak, P.~J. Kenny, R.~Dehak, P.~Dumouchel, and P.~Ouellet, ``Front-end factor analysis for speaker verification,'' \emph{IEEE Transactions on Audio, Speech, and Language Processing}, vol.~19, no.~4, pp. 788--798, 2010.

\bibitem{wan2018generalized}
L.~Wan, Q.~Wang, A.~Papir, and I.~Lopez~Moreno, ``Generalized end-to-end loss for speaker verification,'' in \emph{International Conference on Acoustics, Speech and Signal Processing (ICASSP)}.\hskip 1em plus 0.5em minus 0.4em\relax IEEE, 2018, pp. 4879--4883.

\bibitem{li2017deep}
C.~Li, X.~Ma, B.~Jiang, X.~Li, X.~Zhang, X.~Liu, Y.~Cao, A.~Kannan, and Z.~Zhu, ``Deep speaker: an end-to-end neural speaker embedding system,'' \emph{arXiv preprint arXiv:1705.02304}, vol. 650, 2017.

\bibitem{snyder2018x}
D.~Snyder, D.~Garcia-Romero, G.~Sell, D.~Povey, and S.~Khudanpur, ``X-vectors: Robust {DNN} embeddings for speaker recognition,'' in \emph{International Conference on Acoustics, Speech and Signal Processing (ICASSP)}.\hskip 1em plus 0.5em minus 0.4em\relax IEEE, 2018, pp. 5329--5333.

\bibitem{pelecanos2022parameter}
J.~Pelecanos, Q.~Wang, Y.~Huang, and I.~L. Moreno, ``Parameter-free attentive scoring for speaker verification,'' in \emph{Odyssey: The Speaker and Language Recognition Workshop}, 2022.

\bibitem{snyder2016deep}
D.~Snyder, P.~Ghahremani, D.~Povey, D.~Garcia-Romero, Y.~Carmiel, and S.~Khudanpur, ``Deep neural network-based speaker embeddings for end-to-end speaker verification,'' in \emph{2016 IEEE Spoken Language Technology Workshop (SLT)}.\hskip 1em plus 0.5em minus 0.4em\relax IEEE, 2016, pp. 165--170.

\bibitem{pelecanos2021dr}
J.~Pelecanos, Q.~Wang, and I.~L. Moreno, ``{Dr-Vectors}: Decision residual networks and an improved loss for speaker recognition,'' in \emph{Proc. Interspeech}, 2021.

\bibitem{prabhavalkar2015automatic}
R.~Prabhavalkar, R.~Alvarez, C.~Parada, P.~Nakkiran, and T.~N. Sainath, ``Automatic gain control and multi-style training for robust small-footprint keyword spotting with deep neural networks,'' in \emph{International Conference on Acoustics, Speech and Signal Processing (ICASSP)}.\hskip 1em plus 0.5em minus 0.4em\relax IEEE, 2015, pp. 4704--4708.

\bibitem{mallick2022matchmaker}
A.~Mallick, K.~Hsieh, B.~Arzani, and G.~Joshi, ``Matchmaker: data drift mitigation in machine learning for large-scale systems,'' \emph{Proceedings of Machine Learning and Systems}, vol.~4, pp. 77--94, 2022.

\bibitem{ratner2019mlsys}
A.~Ratner, D.~Alistarh, G.~Alonso, D.~G. Andersen, P.~Bailis, S.~Bird, N.~Carlini, B.~Catanzaro, J.~Chayes, E.~Chung \emph{et~al.}, ``{MLSys}: The new frontier of machine learning systems,'' \emph{arXiv preprint arXiv:1904.03257}, 2019.

\bibitem{timur2019deploying}
T.~D. Timur, I.~K.~E. Purnama, and S.~M.~S. Nugroho, ``Deploying scalable face recognition pipeline using distributed microservices,'' in \emph{International Conference on Computer Engineering, Network, and Intelligent Multimedia (CENIM)}.\hskip 1em plus 0.5em minus 0.4em\relax IEEE, 2019, pp. 1--5.

\bibitem{bolme2020face}
D.~S. Bolme, N.~Srinivas, J.~Brogan, and D.~Cornett, ``Face recognition oak ridge (faro): A framework for distributed and scalable biometrics applications,'' in \emph{International Joint Conference on Biometrics (IJCB)}.\hskip 1em plus 0.5em minus 0.4em\relax IEEE, 2020, pp. 1--8.

\bibitem{limoncelli2014practice}
T.~A. Limoncelli, S.~R. Chalup, and C.~J. Hogan, \emph{The Practice of Cloud System Administration: {DevOps} and {SRE} Practices for Web Services, Volume 2}.\hskip 1em plus 0.5em minus 0.4em\relax Addison-Wesley Professional, 2014, vol.~2.

\bibitem{sanders2019sequential}
P.~Sanders, K.~Mehlhorn, M.~Dietzfelbinger, and R.~Dementiev, \emph{Sequential and Parallel Algorithms and Data Structures}.\hskip 1em plus 0.5em minus 0.4em\relax Springer, 2019.

\bibitem{casanova:hal-01017319}
H.~Casanova, A.~Giersch, A.~Legrand, M.~Quinson, and F.~Suter, ``Versatile, scalable, and accurate simulation of distributed applications and platforms,'' \emph{Journal of Parallel and Distributed Computing}, vol.~74, no.~10, pp. 2899--2917, Jun. 2014.

\bibitem{zheng2004bigsim}
G.~Kakulapati, ``{BigSim}: A parallel simulator for performance prediction of extremely large parallel machines,'' in \emph{International Parallel and Distributed Processing Symposium}.\hskip 1em plus 0.5em minus 0.4em\relax IEEE, 2004, p.~78.

\bibitem{buyya2009modeling}
R.~Buyya, R.~Ranjan, and R.~N. Calheiros, ``Modeling and simulation of scalable cloud computing environments and the cloudsim toolkit: Challenges and opportunities,'' in \emph{International conference on high performance computing \& simulation}.\hskip 1em plus 0.5em minus 0.4em\relax IEEE, 2009, pp. 1--11.

\bibitem{Wang2019}
Q.~Wang, H.~Muckenhirn, K.~Wilson, P.~Sridhar, Z.~Wu, J.~R. Hershey, R.~A. Saurous, R.~J. Weiss, Y.~Jia, and I.~L. Moreno, ``{VoiceFilter}: Targeted voice separation by speaker-conditioned spectrogram masking,'' in \emph{Proc. Interspeech}, 2019, pp. 2728--2732.

\bibitem{Wang2020}
Q.~Wang, I.~L. Moreno, M.~Saglam, K.~Wilson, A.~Chiao, R.~Liu, Y.~He, W.~Li, J.~Pelecanos, M.~Nika, and A.~Gruenstein, ``{VoiceFilter-Lite}: Streaming targeted voice separation for on-device speech recognition,'' in \emph{Proc. Interspeech}, 2020, pp. 2677--2681.

\bibitem{rikhye2021multiuser}
R.~Rikhye, Q.~Wang, Q.~Liang, Y.~He, and I.~McGraw, ``Multi-user {VoiceFilter-Lite} via attentive speaker embedding,'' in \emph{Automatic Speech Recognition and Understanding Workshop (ASRU)}.\hskip 1em plus 0.5em minus 0.4em\relax IEEE, 2021.

\bibitem{rikhye2022closing}
------, ``Closing the gap between single-user and multi-user voicefilter-lite,'' in \emph{Odyssey: The Speaker and Language Recognition Workshop}, 2022.

\bibitem{o2023conditional}
T.~O’Malley, S.~Ding, A.~Narayanan, Q.~Wang, R.~Rikhye, Q.~Liang, Y.~He, and I.~McGraw, ``Conditional conformer: Improving speaker modulation for single and multi-user speech enhancement,'' in \emph{International Conference on Acoustics, Speech and Signal Processing (ICASSP)}.\hskip 1em plus 0.5em minus 0.4em\relax IEEE, 2023, pp. 1--5.

\bibitem{ding2019personal}
S.~Ding, Q.~Wang, S.-y. Chang, L.~Wan, and I.~L. Moreno, ``{Personal VAD}: Speaker-conditioned voice activity detection,'' in \emph{Odyssey: The Speaker and Language Recognition Workshop}, 2020.

\bibitem{ding2022personal}
S.~Ding, R.~Rikhye, Q.~Liang, Y.~He, Q.~Wang, A.~Narayanan, T.~O'Malley, and I.~McGraw, ``Personal vad 2.0: Optimizing personal voice activity detection for on-device speech recognition,'' in \emph{Proc. Interspeech}, 2022.

\bibitem{jia2018transfer}
Y.~Jia, Y.~Zhang, R.~Weiss, Q.~Wang, J.~Shen, F.~Ren, P.~Nguyen, R.~Pang, I.~Lopez~Moreno, Y.~Wu \emph{et~al.}, ``Transfer learning from speaker verification to multispeaker text-to-speech synthesis,'' \emph{Advances in neural information processing systems}, vol.~31, 2018.

\bibitem{alvarez2016efficient}
R.~Alvarez, R.~Prabhavalkar, and A.~Bakhtin, ``On the efficient representation and execution of deep acoustic models,'' in \emph{Proc. Interspeech}, 2016, pp. 2746--2750.

\bibitem{shangguan2019optimizing}
Y.~Shangguan, J.~Li, Q.~Liang, R.~Alvarez, and I.~McGrawn, ``Optimizing speech recognition for the edge,'' \emph{arXiv preprint arXiv:1909.12408}, 2019.

\bibitem{ding20224}
S.~Ding, P.~Meadowlark, Y.~He, L.~Lew, S.~Agrawal, and O.~Rybakov, ``4-bit conformer with native quantization aware training for speech recognition,'' in \emph{Proc. Interspeech}, 2022.

\bibitem{rybakov20232}
O.~Rybakov, P.~Meadowlark, S.~Ding, D.~Qiu, J.~Li, D.~Rim, and Y.~He, ``2-bit conformer quantization for automatic speech recognition,'' \emph{arXiv preprint arXiv:2305.16619}, 2023.

\bibitem{nakkiran2015compressing}
P.~Nakkiran, R.~Alvarez, R.~Prabhavalkar, and C.~Parada, ``Compressing deep neural networks using a rank-constrained topology,'' 2015.

\bibitem{lecun1990optimal}
Y.~LeCun, J.~S. Denker, and S.~A. Solla, ``Optimal brain damage,'' in \emph{Advances in Neural Information Processing Systems}, 1990, pp. 598--605.

\bibitem{wu2021dynamic}
Z.~Wu, D.~Zhao, Q.~Liang, J.~Yu, A.~Gulati, and R.~Pang, ``Dynamic sparsity neural networks for automatic speech recognition,'' in \emph{International Conference on Acoustics, Speech and Signal Processing (ICASSP)}.\hskip 1em plus 0.5em minus 0.4em\relax IEEE, 2021, pp. 6014--6018.

\bibitem{hinton2015distilling}
G.~Hinton, O.~Vinyals, and J.~Dean, ``Distilling the knowledge in a neural network,'' \emph{arXiv preprint arXiv:1503.02531}, 2015.

\bibitem{gupta2021google}
M.~Gupta, ``{Google Tensor} is a milestone for machine learning,'' \emph{Google Pixel Blog}, 2021.

\bibitem{de2017europe}
S.~De~Silva, A.~Liu, and L.~Nabarro, ``Europe's tough new law on biometrics,'' \emph{Biometric Technology Today}, vol. 2017, no.~2, pp. 5--7, 2017.

\bibitem{xia2022turn}
W.~Xia, H.~Lu, Q.~Wang, A.~Tripathi, Y.~Huang, I.~L. Moreno, and H.~Sak, ``{Turn-to-Diarize}: Online speaker diarization constrained by transformer transducer speaker turn detection,'' in \emph{International Conference on Acoustics, Speech and Signal Processing (ICASSP)}.\hskip 1em plus 0.5em minus 0.4em\relax IEEE, 2022, pp. 8077--8081.

\bibitem{ghomi2017load}
E.~J. Ghomi, A.~M. Rahmani, and N.~N. Qader, ``Load-balancing algorithms in cloud computing: A survey,'' \emph{Journal of Network and Computer Applications}, vol.~88, pp. 50--71, 2017.

\bibitem{waskom2021seaborn}
M.~L. Waskom, ``Seaborn: statistical data visualization,'' \emph{Journal of Open Source Software}, vol.~6, no.~60, p. 3021, 2021.

\bibitem{scott2015multivariate}
D.~W. Scott, \emph{Multivariate density estimation: theory, practice, and visualization}.\hskip 1em plus 0.5em minus 0.4em\relax John Wiley \& Sons, 2015.

\bibitem{alakeel2010guide}
A.~M. Alakeel \emph{et~al.}, ``A guide to dynamic load balancing in distributed computer systems,'' \emph{International journal of computer science and information security}, vol.~10, no.~6, pp. 153--160, 2010.

\end{thebibliography}

\end{document}